\documentclass[runningheads]{llncs}
\usepackage{tikz}
\usepackage{amsmath}
\usepackage{url}
\usepackage{subfig}
\usepackage{graphicx,dblfloatfix}
\usepackage{caption}
\usepackage{color}
\usepackage{hyperref}
\usepackage{amsmath}
\usepackage{algorithm}
\usepackage{algorithmic} 
\usepackage{enumerate}
\usepackage{tabularx}
\usepackage{bm}
\usepackage{multirow,color}
\usepackage{threeparttable}
\usepackage{setspace}

\newcommand*\circled[1]{\tikz[baseline=(char.base)]{
            \node[shape=circle,draw,inner sep=1pt] (char) {#1};}}
            
\begin{document}

\title{Fast and Efficient Malware Detection with Joint Static and Dynamic Features Through Transfer Learning}
\titlerunning{Malware Detection with Static-Dynamic Features using Transfer Learning}

\author{Mao~V.~Ngo\inst{1} \and Tram Truong-Huu\inst{2} \and Dima Rabadi\inst{3} \and Jia Yi Loo\inst{4} \and Sin G. Teo\inst{4}}

\institute{Singapore University of Technology and Design, Singapore 
           \email{vanmao\_ngo@sutd.edu.sg}\\ \and
           Singapore Institute of Technology, Singapore 
           \email{truonghuu.tram@singaporetech.edu.sg} \\ \and
           Penn State Shenango, Pennsylvania, USA\\ 
           \email{dqr5554@psu.edu}\\\and
            Institute for Infocomm Research, A*STAR, Singapore
            \email{\{loojy,teosg\}@i2r.a-star.edu.sg}}

\maketitle

\begin{abstract}
In malware detection, dynamic analysis extracts the runtime behavior of malware samples in a controlled environment and static analysis extracts features using reverse engineering tools. While the former faces the challenges of anti-virtualization and evasive behavior of malware samples, the latter faces the challenges of code obfuscation. To tackle these drawbacks, prior works proposed to develop detection models by aggregating dynamic and static features, thus leveraging the advantages of both approaches. However, simply concatenating dynamic and static features raises an issue of imbalanced contribution due to the heterogeneous dimensions of feature vectors to the performance of malware detection models. Yet, dynamic analysis is a time-consuming task and requires a secure environment, leading to detection delays and high costs for maintaining the analysis infrastructure. In this paper, we first introduce a method of constructing aggregated features via concatenating latent features learned through deep learning with \textit{equally-contributed dimensions}. We then develop a \textit{knowledge distillation} technique to transfer knowledge learned from aggregated features by a teacher model to a student model trained only on static features and use the trained student model for the detection of new malware samples. We carry out extensive experiments with a dataset of $86\,709$ samples including both benign and malware samples. The experimental results show that the teacher model trained on aggregated features constructed by our method outperforms the state-of-the-art models with an improvement of up to $2.38\%$ in detection accuracy. The distilled student model not only achieves high performance ($97.81\%$ in terms of accuracy) as that of the teacher model but also significantly reduces the detection time (from $70\,046.6$ ms to $194.9$ ms) without requiring dynamic analysis. 
\end{abstract}


\section{Introduction} 

Malicious software, also known as \textit{malware}, is targeted to steal information of computer users, spread the virus into the computer networks, encrypt data for ransom, or do other nefarious purposes. Detecting malware that could be legitimately downloaded by authorized users is a primitive feature of a reliable computer system. To this end, malware detection models using machine learning or deep learning have been proposed to classify samples as malicious or benign. The suspicious samples are analyzed by malware analysis techniques to extract the features used by the detection model.

Malware analysis techniques are categorized into two types: \textit{static} analysis and \textit{dynamic} analysis. Static analysis extracts features of examined samples by dissembling the samples using reverse engineering tools and looking for specific strings or patterns in the binary code; whereas dynamic analysis detonates the samples in an isolated (and secure) environment (i.e., sandboxes) to obtain runtime behavior. While static analysis is fast to collect features, it cannot deal with code obfuscation (e.g., polymorphism, metamorphism, or compression) techniques~\cite{staticAnalysisLimit_2007,DynamicAnalysisSurvey_ACM_Survey2019}. On the other hand, dynamic analysis is more robust against code obfuscation but it requires executing the samples in a sandbox, which is a time-consuming task~\cite{rhode2018early}. In addition, dynamic analysis cannot detect malware equipped with evasive techniques such as anti-virtualization or time delay.

Existing works~\cite{Damodaran2017comparison,han2019maldae,Dima_ACSAC2020,SANTOS2013_OPCODE} use either dynamic features\footnote{Dynamic/static features are features extracted from dynamic/static analysis method.}~\cite{kadiyala2020program,Dima_ACSAC2020,rhode2018early}, static features~\cite{Anderson2018Ember,SANTOS2013_OPCODE,partha:2021}, or aggregation of static and dynamic features~\cite{han2019maldae} to train a detection model. As presented in~\cite{han2019maldae}, the model trained with aggregated features can generally boost detection performance as it overcomes the drawbacks of static analysis and dynamic analysis. However, the selection of static and dynamic feature sets for aggregation is a crucial task as it can positively or negatively affect the model's performance. Furthermore, our preliminary experiments also show that naively concatenating static and dynamic feature vectors with heterogeneous dimensions faces an issue of the imbalanced contribution of the feature vectors to the model's performance. The longer feature vector may dominantly contribute to the model's performance while the shorter feature vector has less impact~\cite{oramas:2017}. This also increases the dimension of the aggregated feature vector, thus subsequently increasing the model size and complexity. The work in~\cite{han2015deepCompression} proposes to prune near-zero weights from large neural networks or quantize the model by using fewer bits for weights and biases but sacrificing the model performance. In this paper, we continue advocating the aggregation of static and dynamic features. We propose a novel aggregation approach that combines the latent representation of dynamic and static feature vectors with an equal size to train a detection model. 

While aggregating dynamic and static features improves the performance of the detection model as both feature vectors complement to overcome the drawbacks of each other, the inference process for a new sample is still time-consuming due to the extraction of runtime behavior through dynamic analysis. To address this challenge, we propose a transfer learning approach using knowledge distillation (KD)~\cite{hinton2015distilling} to transfer knowledge from a large \textit{teacher} model trained with aggregated features to a small \textit{student} model trained with only static features. The training of the student model aims to minimize a distillation loss function that is a weighted combination of cross-entropy loss and Kullback-Leibler (KL) divergence loss. Consequently, the trained student model enables a fast and efficient inference as it can leverage both rich features from aggregated features (i.e., dynamic and static features) while it does not require malware samples to be analyzed in sandboxes. This significantly reduces detection time and addresses the privacy issues of the sandboxes deployed in clouds. The trained student model can be deployed on commodity computers to perform malware detection and protect them from attacks.   

In summary, we make the following contributions to this work:
\begin{itemize}
    
    \item We propose a method of feature aggregation by using a deep learning model to learn the latent representations of static and dynamic feature vectors. The latent representations will have the same size, thus equally contributing to the malware detection model. 
    
    \item We develop novel one-dimensional convolutional neural network (1D-CNN) architectures, each being used to train a detection model for individual static and dynamic feature vectors and the aggregated feature vector.
    
    \item We develop a knowledge distillation technique to transfer rich knowledge from a large teacher model trained with the aggregated feature vector to a small student model trained only with static features. 
    
    \item We carry out extensive experiments to evaluate the proposed feature aggregation approach and the performance of the student model obtained from the knowledge distillation.  
\end{itemize}

The rest of the paper is organized as follows. Section~\ref{sec:relatedwork} discusses the related work. Section~\ref{sec:MalwareAnalysis} presents the background of malware analysis and feature extraction. Section~\ref{sec:dlmodels} presents deep learning-based malware detection models. Section~\ref{sec:transferlearning} presents our proposed transfer learning approach for fast and efficient malware detection. Section~\ref{sec:performanceEvaluation} presents extensive experiments to evaluate the performance of the proposed approach. Section~\ref{sec:conclusion} concludes the paper.

\section{Related Work}
\label{sec:relatedwork}

\subsection{Static Malware Analysis}
Static analysis is used to extract syntactical information (referred to as static features) from the binary Windows portable executable (PE) files such as imported and exported functions, base addresses of the section headers, debug information, or even operational codes (via disassembly tools such as Ghidra~\cite{Ghidra} or IDA Pro~\cite{IDA_Pro}). Operational code (opcode)-based static features have been widely used in literature~\cite{SANTOS2013_OPCODE,fan2016malicious,nOpcode_Kang2016,zhang2019feature}. In~\cite{fan2016malicious}, the authors proposed to use the frequency of top $k$ sequences of the $n$-gram opcodes to reduce the useless information caused by undiscriminating instructions. Similarly, in~\cite{nOpcode_Kang2016}, the authors proposed two opcode-based feature vectors used for malware detection: the frequency (i.e., the number of occurrences) and binary (i.e., whether the $n$-gram opcodes exist or not). In~\cite{zhang2019feature}, the authors implemented a CNN model for malware detection and type classification (e.g., Backdoor, Trojan, Worm, etc.) using static features. The authors disassembled Windows binaries using the W32dasm~\cite{W32DASM} tool and integrated two extracted static features: API call frequency and opcode bi-gram. Such works that rely on only static features are vulnerable to obfuscation techniques and packing, which constructs samples to be hard to disassemble or extract their opcodes (anti-reverse engineered).

\subsection{Dynamic Malware Analysis} 

The dynamic analysis uses a virtual environment (e.g., virtual machines and sandboxes) to collect malware runtime behavior. Based on dynamic analysis reports, specific features are extracted to build detection models to flag suspicious behavior such as an unusual pattern of API/system calls, or calls to the attacker's command-and-control (CnC) servers. Recently, Zhang \textit{et al.}~\cite{Zhang2020ScarecrowDE} proposed a deceptive engine called Scarecrow to exploit evasion techniques by transforming or camouflaging the physical end-user environment into an analysis-like environment from the view of evasive malware. But the battle between malware authors and defenders is never-ending. Novel evasive techniques exploit an observation that files are detonated in sandboxes emulating physical machines but without human interaction. So attackers develop malware that remains passive until it detects signs of human users, e.g., a mouse click, intelligent response to dialog boxes, or the user scrolls to the second page of a Rich Text Format (RTF) document~\cite{EvasiveBehaviors_fireeye2013}. Another tactic used by attackers is simply to delay the execution of the malicious code by adding extended sleep calls as malware samples are detonated in automatic sandbox environments within a short period (e.g., $2$ minutes by default in the Cuckoo sandbox).
Many reported malware (e.g., Trojan Nap--February $2013$, Kelihos Botnet--$2011$) refrain from their suspicious behavior through a monitoring process (e.g., $10$ minutes in Nap Trojan code).  

\subsection{AI-based Malware Detection}

Static and dynamic features have their advantages and drawbacks. Thus, using heterogeneous features would help achieve harder-to-be-bypassed detection models. In~\cite{Damodaran2017comparison}, the authors compared the performance of HMM model on dynamic, static, and hybrid (train on dynamic and test on static) feature sets. Their results show the best performance obtained with the dynamic feature set. In~\cite{ndibanje2019cross}, the authors developed a hybrid framework to classify Windows samples into benign and malicious using LCS and cosine similarity-based machine learning algorithms. The algorithms are trained on combined static and dynamic API sequences to learn malicious and benign behavior. In~\cite{han2019maldae}, the authors developed MalDAE that analyzes the correlation and semantic mapping between static and dynamic API sequences. The hybrid API sequences of samples are used to train machine learning models to detect malicious samples. Compared to our proposed model, the above works rely only on the sequence of the API calls and ignore the API arguments. In~\cite{9322377}, the authors proposed to convert binary files to images with pre-defined sizes and used a convolution neural network for malware detection. This approach is similar to static analysis as malware signatures create byte sequences that differentiate each other.

\subsection{Transfer Learning for Malware Detection} 

Recently, Zhi \textit{et al.}~\cite{zhi2021lightweightKD} used knowledge distillation (KD) to transfer knowledge from a large teacher model to a lightweight student model, both using dynamic features for malicious Android app detection. The student model still depends on dynamic analysis, which incurs long delays due to the execution of samples in an Android emulator environment to extract runtime behavior during the detection phase. In contrast, our lightweight student model uses only static features, which are fast to obtain and can be easily deployed on commodity computers. 

\section{Malware Analysis and Feature Extraction}
\label{sec:MalwareAnalysis}

In this section, we briefly introduce our static and dynamic malware analysis for feature extraction.

\subsection{Static Analysis and Feature Extraction}

We use two feature extraction methods to extract two feature vectors, namely EMBER and OPCODE (OPerational CODE). 

\paragraph{\bf EMBER features} We use the Library to Instrument Executable Formats (LIEF)~\cite{LIEF} to parse PE files, then extract EMBER (Endgame Malware BEnchmark for Research) features~\cite{Anderson2018Ember}. Specifically, the EMBER feature vector consists of eight groups: five groups of parsed features (e.g., general file information, header information, imported and exported functions, and section information), and three groups of format-agnostic features (e.g., byte histogram, byte entropy histogram, and string extraction). Eventually, we obtain a feature vector whose size is $2381$ for each malware sample. 
    
\paragraph{\bf OPCODE features} We use a free reverse engineering tool, Ghidra~\cite{Ghidra}, to perform static analysis on binary PE files, resulting in an output report containing the OPCODE sequence. We construct $3$-gram OPCODEs, each being a sequence of 3 consecutive OPCODEs and considered as a feature in the OPCODE feature vector. As the total number of $3$-gram OPCODE is large (more than a million in our experiments), we first conduct feature reduction by excluding one near-identity dummy features that have Pearson correlations close to $1$ (e.g., above a threshold of $0.95$). And then, we do feature selection based on Mutual Information (MI), which is also used for feature selection of $n$-gram OPCODE in Android malware detection~\cite{nOpcode_Kang2016}. Specifically, we calculate MI of each $3$-gram OPCODE feature to a classification class (i.e., benign or malicious) to select the top $3$-gram OPCODEs (e.g., above $98$\textsuperscript{th} percentile). We empirically obtain an OPCODE feature vector whose size is $33\,338$ for training malware detection models and prediction of new samples.

\subsection{Dynamic Analysis and Feature Extraction} 
We adopt the API-ARG feature extraction method developed in~\cite{Dima_ACSAC2020}\footnote{In~\cite{Dima_ACSAC2020}, API-ARG was reported with a different name \texttt{Method2}.} to construct the dynamic feature vector. Here, we briefly describe the method for the sake of completeness. We use a Cuckoo sandbox host deploying with a Windows-7 OS on a virtual machine to capture the runtime behavior of malware samples. Based on the behavioral analysis JSON reports from the Cuckoo sandbox, we extract sequences of both Application Programming Interfaces (API) and their arguments. Suffixes of API names  (e.g., \texttt{Ex, A, W, ExA, ExW}) are removed and similar API calls will be merged, e.g., \texttt{FindFirstFileExW} and \texttt{FindFirstFileExA} are merged into \texttt{FindFirstFile}. After pre-processing API arguments (e.g., convert integer argument into logarithmic bins, categorize special directory/path/url such as \texttt{system32}, specify if special registry keys or special commands are available), the API name is concatenated with each of its arguments to form separate string features. As a result, the number of features is a product of the number of API names and the number of their arguments. Finally, a Hashing Vectorizer function~\cite{HashingVectorizer} is used to encode these string features into a binary vector, whose length is $2^{20}=1\,048\,576$.

\subsection{Feature Aggregation}
We combine the static feature (i.e., EMBER and OPCODE), and the dynamic feature (API-ARG) to construct an aggregated feature vector as another feature engineering technique. The intuition of combining both the static and dynamic features is to obtain complimentary advantages of each type of feature. Suppose we aggregate a static feature vector $\mathbf{X}_\text{static}$ with input dimension $d_s$ and a dynamic feature vector $\mathbf{X}_\text{dynamic}$ with input dimension $d_d$, intuitively, we can either (i) combine them in a weighted sum with padding zeros for a smaller dimensional vector, or (ii) concatenate them to produce a larger vector with dimension $d_s+d_d$. However, the above approaches face an issue of imbalanced contribution from individual feature vectors, especially when one vector is dominantly larger than the other (e.g., the API-ARG vector with $1\,048\,576$ dimensions is much larger than the EMBER vector with $2381$ dimensions). 

In order to tackle the imbalanced contribution issue, instead of directly using raw feature vectors, we propose to extract and compress each raw feature vector into a smaller dimensional space of the same size. This can be done using a deep neural network as a feature extractor. With equal-sized latent representations, we can eliminate the imbalanced contribution issue when aggregating static and dynamic feature vectors. Subsequently, we can either (i) concatenate the latent representation vectors forming a longer vector or (ii) combine these vectors via a weighted sum manner to the same dimensional vector. In this work, we choose the first option and then feed the aggregated vector into a fully connected (FC) layer for classification. The rationale behind this is that weighted parameters of the FC layer can be interpreted as tunable weights of the weighted sum of not only two representation vectors but also within different features in the projected vector. So concatenating two (or more) latent representation vectors and then feeding them into an FC layer is considered a comprehensive way to aggregate the latent representation vectors. We provide further details on the architectures of deep neural networks in Section~\ref{subsec:ArchitectureAgg}.

\section{Deep Learning-based Malware Detection Models}
\label{sec:dlmodels}

\begin{figure}[t]
    \centering
    \includegraphics[width=.95\linewidth]{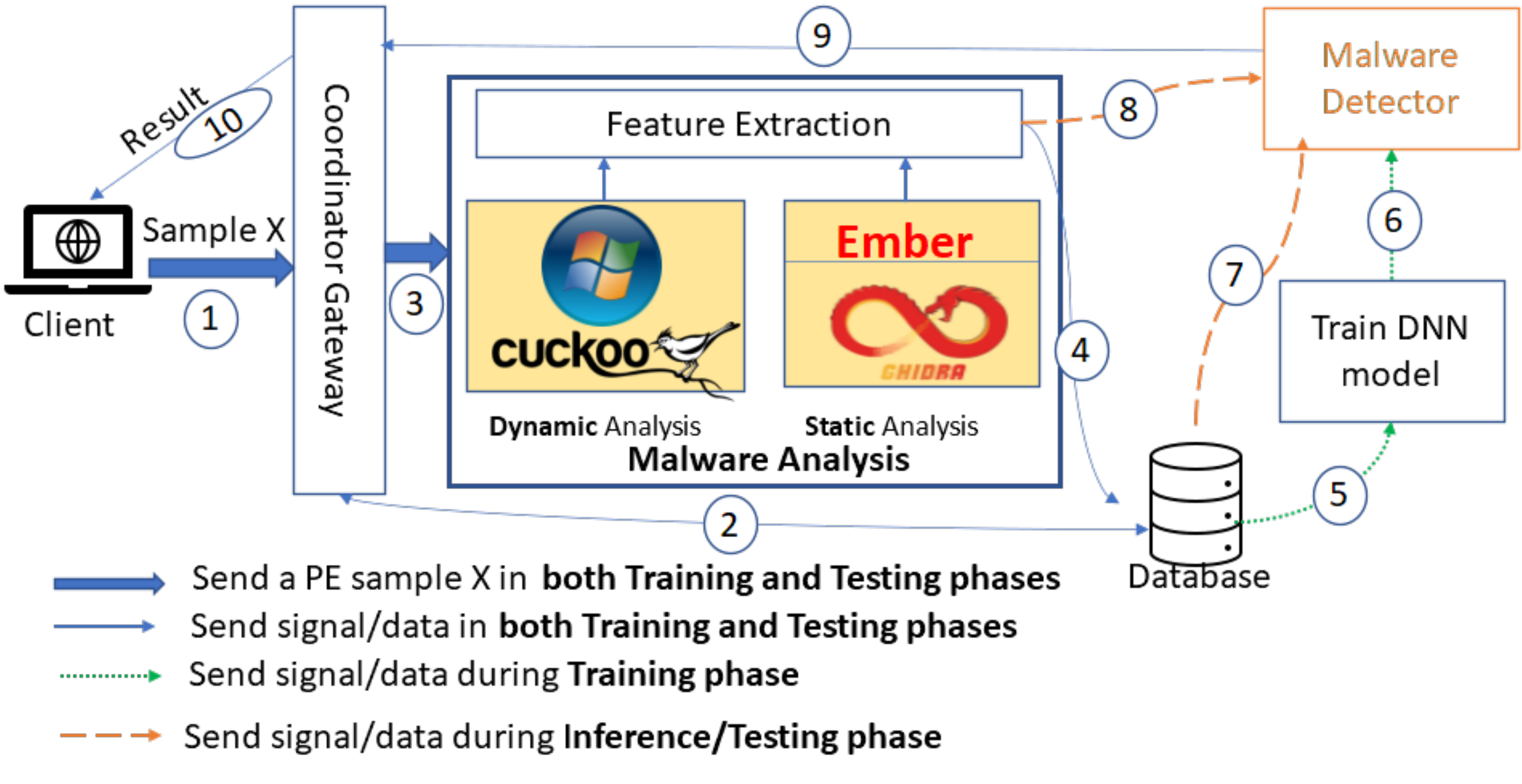}
    \caption{Deep learning-based malware detection system.}
    \label{fig:malwareDetectionFlow}
    \vspace{-2.5ex}
\end{figure}

Fig.~\ref{fig:malwareDetectionFlow} shows workflows of training and testing/deploying stages in our deep learning (DL)-based malware detection system. 
First, a \textit{Client} sends a portable executable (PE) sample $X$ to a \textit{Coordinator Gateway} for virus-scanning, as shown in step \circled{1}. 
The Coordinator Gateway calculates the hash value of the PE file, and checks the availability of its dynamic and static features in a \textit{Database}, as shown in step \circled{2}. If they are available, the extracted features of the sample are retrieved, and proceed in step \circled{5} to the model training phase or step \circled{7} in the  testing phase. If the PE file is new to the system, will be analyzed in the \textit{Malware Analysis} module (step \circled{3}) (including \textit{Dynamic Analysis} and \textit{Static Analysis}), and output reports are passed through the  \textit{Feature Extraction} module to obtain dynamic features and static features (which have been presented in Section~\ref{sec:MalwareAnalysis}). These features are stored in the Database as shown in step \circled{4}, and sent to either: the \textit{Malware Detector} module via step \circled{8} if the system is in the inference stage, or the \textit{Train DNN model} module via step \circled{7} if the system is in the training stage. The trained model is deployed for inference in step \circled{6}. In the inference stage, the detection result is sent back to the client via the Coordinator Gateway, as shown in lines \circled{9} and \circled{10}. 

The \textit{Train DNN model} module trains a DL model for a binary classification task (i.e., benign or malicious) in a supervised learning manner. Instead of using a single deep neural network trained on a particular feature vector, we develop multiple neural network architectures, each being used for a feature vector either static, dynamic, or aggregated feature vectors. Below, we present DL-based network architectures for various feature vectors. 

\subsection{Neural Network Architectures for Individual Feature Vectors}
\label{subsec:ArchitectureSingleFeature}

\begin{figure}[t]
    \centering
    \includegraphics[width=0.8\linewidth]{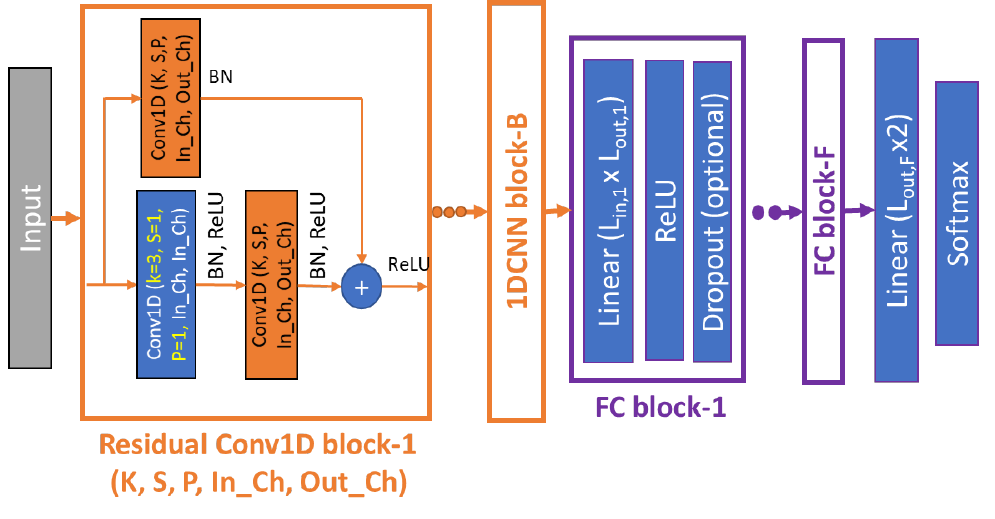}
    \caption{Architecture of 1D-CNN with Residual Blocks.}
    \label{fig:Net5_architecture}
    \vspace{-2ex}
\end{figure}

Inspired by the success of residual deep neural networks~\cite{he2016deep_ResNet} in computer vision, we design three one-dimensional residual deep neural networks, namely ResNet1D for learning the latent representation of individual feature vectors (i.e., EMBER, OPCODE, and API-ARG). Unlike images that have a two-dimensional input, malware features are represented as one-dimensional vectors. Therefore, we design our ResNet1D with a high stride step (up to $8$ steps) within each ResNet1D-block. In Fig.~\ref{fig:Net5_architecture}, the Residual Conv1D block-1 shows the details of the ResNet1D-block that consists of two paths: (i) the bottom path is the main learning feature path, and (ii) the top path is a residual shortcut path. This design is similar to the \textit{bottleneck} block~\cite{ResNetImplementation} of ResNet. Due to the size of the model and compressing ratio design for each layer, we implement a single ResNet1D block for each layer to keep the model small while obtaining good performance. Detailed neural network architectures for the EMBER feature vector, OPCODE feature vector, and API-ARG feature vector are shown in Table~\ref{tab:detailedNetworkArchitectEmber}, Table~\ref{tab:detailedNetworkArchitectOpcode}, and Table~\ref{tab:detailedNetworkArchitectAPIARG} (in Appendix~\ref{sec:AppArchtiectureIndividualFeatures}), respectively. For all three networks, we consider an expected number of layers for the whole 1D-CNN model and the size of each individual feature vector. We design stride steps and kernel sizes of these layers to incrementally compress input dimensions and end up with 1-dimensional output; while the padding (either $1$ or $0$) is added in order to make the last sliding window well fit the kernel size. Together with compression of the input dimensions, we expand the model in width by increasing the number of output channels, i.e., mostly double after each layer. These architectures were obtained after extensive experiments for parameter optimization.

Besides the ResNet1D block, we also consider other recent advanced variants of ResNet in the computer vision domain such as ResNeXt~\cite{xie2017ResNeXt}, inverted ResNeXt~\cite{xie2017ResNeXt}, and ConvNeXt~\cite{liu2022convnet}. We adopt key techniques of these variants used with conv-2D, implement them to conv-1D and adjust the ResNet1D block to construct ResNeXt-1D, Inverted RestNeXt-1D, and ConvNeXt-1D blocks, which are presented in Appendix~\ref{sec:ArchitectureAggRawFeatures}.
Since we only replace the basic block architecture, we reuse the high-level network architecture as presented in Table~\ref{tab:detailedNetworkArchitectEmber}, Table~\ref{tab:detailedNetworkArchitectOpcode}, and Table~\ref{tab:detailedNetworkArchitectAPIARG} for each individual feature vector. We changed the first layer of Conv1D with $\texttt{kernel}=3$ and $\texttt{padding}=1$ instead of $\texttt{kernel}=1$ and $\texttt{padding}=0$ after extensive ablation studies (Section~\ref{subsec:ablationNetwork}).

\subsection{Neural Network Architecture for Aggregated Feature Vectors}
\label{subsec:ArchitectureAgg}

To construct the aggregated feature vector, besides the naive approach of concatenating original feature vectors into a long vector, we introduce a novel approach of using latent presentation learned by a 1D-CNN extractor.

\paragraph{\bf Concatenate original feature vectors} We can concatenate two static original feature vectors (EMBER with $2381$ dimensions) and OPCODE with $33\,338$ dimensions) and dynamic original feature vector (API-ARG with $1\,048\,576$ dimensions). However, this naive aggregation method yields an imbalanced contribution from three individual feature vectors. The largest feature vector---API-ARG---dominantly affects the aggregated feature vector. In Appendix~\ref{sec:ArchitectureAggRawFeatures}, we present 1D-CNN architectures for concatenating $2$ original feature vectors (EMBER + API-ARG) and $3$ original feature vectors (EMBER + OPCODE + API-ARG) in Table~\ref{tab:Agg2RawNet} and Table~\ref{tab:Agg3RawNet}, respectively.
   
\begin{figure}[t]
    \centering
    \includegraphics[width=0.8\linewidth]{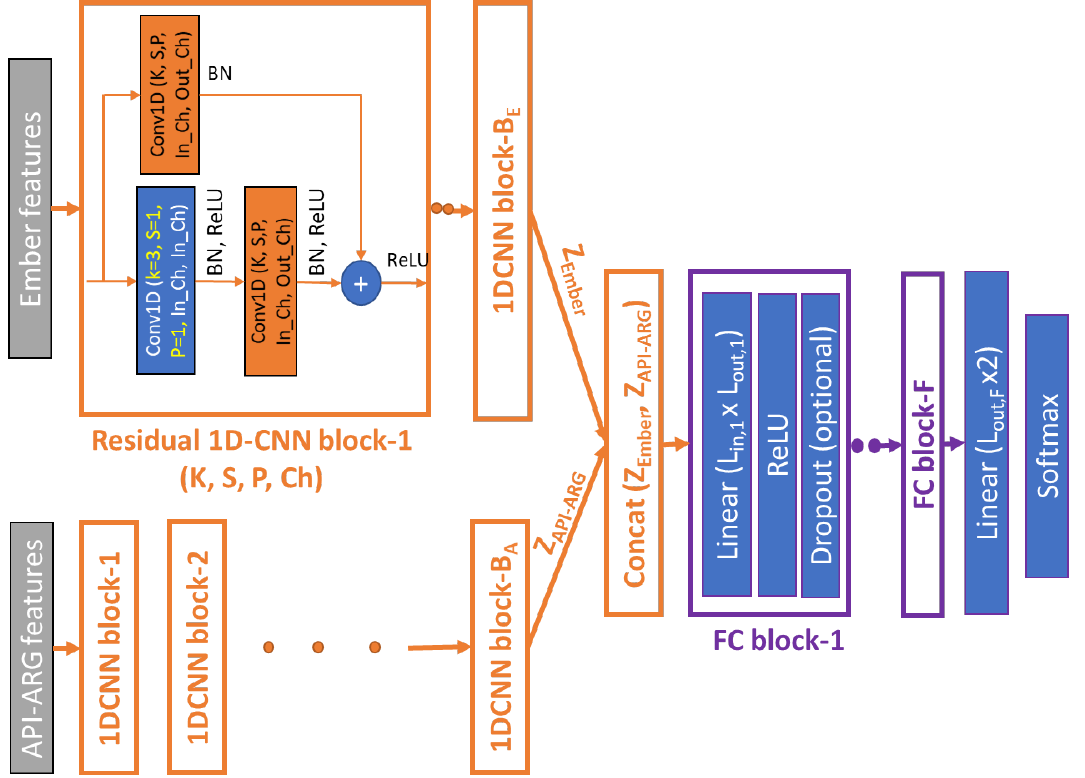}
    \caption{Architecture of 1D-CNN for Learning Latent Representation and a Feed Forward Network for Classification.}
    \label{fig:featNet5_AggregatedFeatureArchitecture}
    \vspace{-2ex}
\end{figure}

\paragraph{\bf Concatenate latent representation vectors} Each individual feature vector is passed through a respective 1D-CNN extractor for learning a latent representation vector, which ends up with $384$ dimensions. Hence, concatenating them will eliminate the imbalanced contribution issue. The aggregated feature vector is then fed into a few layers of a feed-forward network to classify samples as either benign or malicious. Figure~\ref{fig:featNet5_AggregatedFeatureArchitecture} shows the architecture of the entire deep learning model including 1D-CNN models for learning latent representation of original feature vectors and a feed-forward network for classification.

While the model trained with the aggregated feature vector leverages the advantages of static and dynamic features, the inference stage is still a time-consuming task for new samples that have not been analyzed in a sandbox to extract dynamic features. To address this challenge, we propose to use knowledge distillation to transfer knowledge from a large \textit{teacher model trained on the aggregated feature vector} to a small \textit{student model trained on a single static feature vector}. In the next section, we present such a knowledge distillation technique.

\section{Transfer Learning with Knowledge Distillation}
\label{sec:transferlearning}

Transfer learning is a technique of training a new model or fine-tuning a pre-trained model to adapt to a new task or a new domain. While transfer learning keeps the network architecture nearly the same or at least the first few feature extraction layers as the original model, the knowledge distillation (KD) approach does not have this constraint. Indeed, knowledge distillation tends to transfer the knowledge from a large \textit{teacher} network to a smaller student network with an aim of approximating the performance of the \textit{student} network as well as that of the teacher network. KD was first introduced by Bucila \textit{et al.}~\cite{Bucila_MC_KDD2006} for model compression and became popular after Hinton \textit{et al.}~\cite{hinton2015distilling} generalized it. With a smaller architecture, the student model typically requires fewer computing resources and enables a fast inference. In this work, we adopt the KD idea to the cybersecurity domain with multiple objectives. We not only transfer knowledge from a large-complex network to a smaller one to achieve the original purpose of KD but we also leverage static and dynamic features of malware analysis to transfer the knowledge of the model trained on aggregated features to a model trained on static features only.

In order to perform KD, we define a joint loss function of both ground-truth labels (i.e., ``hard'' labels) and ``soft'' labels obtained from the distribution of class probabilities predicted by the teacher model.
Specifically, the soft labels are the output of a \texttt{softmax} function of the teacher model's logits.  
However, in many cases, the probability distribution outputs the correct class with a very high probability and very close to zero for other classes. So, it does not provide much information beyond the ground-truth labels (i.e., hard labels). To address this issue, Hinton \textit{et al.}~\cite{hinton2015distilling} proposed the concept of ``softmax temperature''. The softened \texttt{softmax} with a scaling temperature $\tau$ of a network $f$, given input $\bm{x}$ and the last output logit $\bm{z}$ has a probability denoted by $p_i^{f}(\tau)$ of class $i$ calculated from the logit $\bm{z}^f$, and it is defined as follows:
\begin{equation}
    p_i^{f}(\tau) = \frac{\exp{(z_i^{f}/\tau)}}{ \sum_j{\exp{(z_j^{f}/\tau)}}}.
    \label{eq:softenedSoftmax}
\end{equation}
When $\tau=1$, we get the standard \texttt{softmax} function. When $\tau$ increases, the probability distribution generated by the \texttt{softmax} function becomes softer, providing more fine-grained information about the similarity between the predictions of the teacher model and the ground-truth labels.

During the training of deep neural networks, we typically minimize cross-entropy loss between the output probabilities (i.e., after \texttt{softmax} function) and the target labels. In knowledge distillation, the objective of training a smaller student model is to minimize a linear combination of two losses: the cross-entropy (CE) loss with hard labels (as usual) and the Kullback-Leibler (KL) divergence loss between softened probability predictions of the student model and the teacher model. The combined loss is computed as follows:
\begin{align}
\label{eq:lossesKL}
    \mathcal{L}_{KD-KL}(\bm{x}; \bm{W}) =  \alpha  \mathcal{L}_{CE}(\bm{p}^{S}(1), \bm{y}) + (1 - \alpha)  \mathcal{L}_{KL}(\bm{p}^{T}(\tau), \bm{p}^{S}(\tau)),
\end{align}
where $\bm{x}$ is the input feature vector, $\bm{W}$ is the parameter set of the student model, $S$ is the student model, $T$ is the teacher model, $\bm{y}$ is the ground-truth label, and $\alpha$ is a hyper-parameter of the linear combination. $\bm{p}^{S}(\tau)$ and $\bm{p}^T(\tau)$ are the softened probabilities with temperature $\tau$ of the student and teacher models, respectively. The cross-entropy loss is defined as in Eq.~\eqref{eq:CELoss}, and the KL-divergence loss is defined as in Eq.~\eqref{eq:KLloss}:
\begin{align}
\mathcal{L}_{CE}(\bm{p}^{S}(1), \bm{y})  &= \sum_i -y_i \log p_i^S(1), \label{eq:CELoss} \\
\mathcal{L}_{KL}(\bm{p}^{T}(\tau), \bm{p}^{S}(\tau))  &= \tau^2 \sum_i p_i^T(\tau) \log \frac{p_i^T(\tau)}{p_i^S(\tau)}. \label{eq:KLloss}
\end{align}
In~\cite{hinton2015distilling}, the standard choice of $\alpha$ is $0.1$ but it was also examined with a value of $0.5$. The scaling temperature $\tau$ was recommended to take a value in the set of $\{3,4,5\}$~\cite{CompareKLMSE_IJCAI2021,hinton2015distilling}.

\begin{figure}[t]
    \centering
    \includegraphics[width=0.8\linewidth]{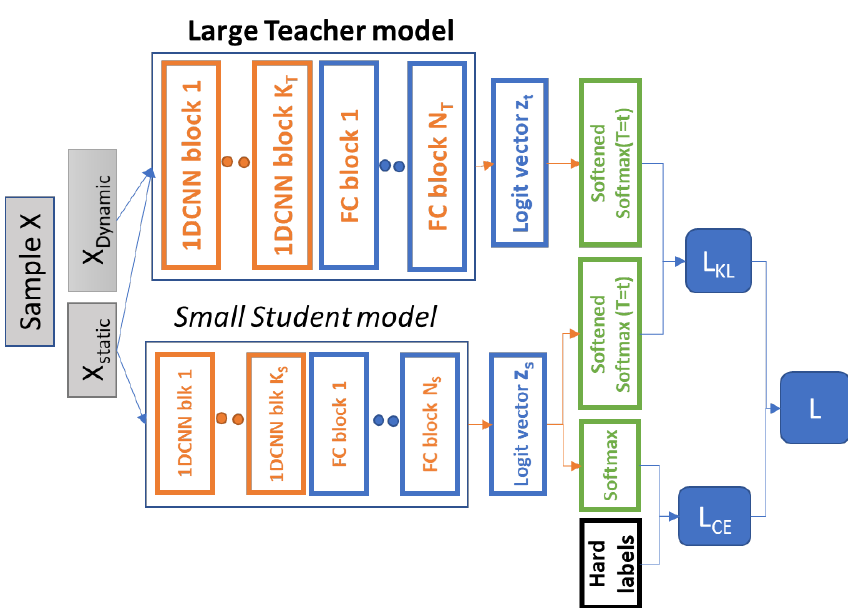}
    \caption{Knowledge Transfer from a Teacher model to a Student model with knowledge distillation (KD).}
    \label{fig:transferLearningDiagram}
    \vspace{-2ex}
\end{figure}

Fig.~\ref{fig:transferLearningDiagram} shows the process of knowledge distillation from a teacher model trained on the aggregated feature vector (e.g., aggregation of  $\mathbf{X}_\text{static}$ and $\mathbf{X}_\text{dynamic}$) to a student model trained with a static feature vector only (i.e.,  $\mathbf{X}_\text{static}$). During the training process, a sample $X$ is passed through the Malware Analysis module (see Fig.~\ref{fig:malwareDetectionFlow}) including dynamic analysis and static analysis. The analysis reports are processed to extract the dynamic and static feature vectors, namely $\mathbf{X}_\text{dynamic}$ and $\mathbf{X}_\text{static}$, respectively. We aggregate dynamic and static feature vectors to train the teacher model (ref. Section~\ref{subsec:ArchitectureAgg}). Subsequently, we use the teacher model to obtain the penultimate output layer (a vector of $384$ dimensions), which is the \texttt{logit} vector $Z^{T}$ in  Fig.~\ref{fig:transferLearningDiagram}.
Finally, we train the student model on the static feature vector $\mathbf{X}_\text{static}$ with a loss function defined as a weighted combination of the CE loss and a KL-divergence loss.

As Kim \textit{et al.}~\cite{CompareKLMSE_IJCAI2021} showed that a large value of $\tau$, strong softening, leads to \textit{logit matching}, whereas a small $\tau$ results in \textit{label matching}. They suggested that ``\texttt{logit} matching has a better generalization capacity than label matching'' when the data has correct labels. As a result, they proposed to use mean-squared error (MSE) loss to directly match the output \texttt{logit} of the teacher model and those of the student model, yielding better performance in most of the cases with correct labels, and remove an unreasonable assumption of zero-mean of the \texttt{logit} generated by the student model~\cite{hinton2015distilling,CompareKLMSE_IJCAI2021}. Furthermore, MSE loss does not require fine-tuning a suitable temperature to smooth the output probabilities. The KD loss function for enhancing the \texttt{logit} matching is then defined as follows:
\begin{align}
\label{eq:lossesMSE}
    \mathcal{L}_{KD-MSE}(\bm{x}; \bm{W}) = \alpha  \mathcal{L}_{CE}(\bm{p}^{S}(1), \bm{y}) + (1-\alpha)  \mathcal{L}_{MSE}(\bm{z}^{T}, \bm{z}^{S}),
\end{align}
where $\mathcal{L}_{MSE}(\bm{z}^{T}, \bm{z}^{S}) = || \bm{z}_S - \bm{z}^T||_2^2$. In this paper, we examine both loss functions, namely KD-KL loss in Eq.~\eqref{eq:lossesKL}, and KD-MSE loss in Eq.~\eqref{eq:lossesMSE} during the knowledge distillation from the teacher model to the student model. 

\paragraph{\bf Maximizing Transferred Knowledge with Ensemble of Multiple Teacher Models}

Due to the randomness during the training process of the teacher model (e.g., using various training datasets, sample shuffle with batches), multiple teacher model instances could have different performances. To maximize the amount of knowledge transferred to the student model, we use an ensemble approach to combine predictions of multiple teacher model instances in an average manner before computing the loss during the training of the student model, thus reducing the variance of the predictions of the teacher models. It is to be noted that the ensemble model increases the inference delay, but it only applies during knowledge distillation (i.e., training of the student model) from the cumbersome ensemble teacher model to the student model.

\section{Performance Evaluation}
\label{sec:performanceEvaluation}

\subsection{Dataset}
\label{subsec:dataset}

We use a private dataset that is bought from a security vendor in 2020. The dataset consists of $86\,709$ Windows portable executable (PE) samples (including $42\,742$ benign and $43\,967$ malicious samples)\footnote{The hash value of samples will be provided on request for experiment reproducibility.}. We split the dataset with an $80$:$20$ ratio into a training set with $69\,367$ samples and a test set with $17\,342$ samples. We then split the training set with an 80:20 ratio into a development set and a validation set to fine-tune the best hyper-parameters (in Section~\ref{subsec:setupHyperparameters}). Subsequently, we use the best hyper-parameters to train the final model on the whole training set, and test on the test set for performance evaluation. We perform static and dynamic analysis as described in Section~\ref{sec:MalwareAnalysis} to extract EMBER features, OPCODE features, and dynamic API-ARG features for each sample. To speed up the dynamic analysis process, we develop a distributed Cuckoo infrastructure that allows concurrent analysis of multiple samples in multiple sandboxes. A detailed description of the infrastructure is given in Appendix~\ref{appendix:dist-cuckoo}.

\subsubsection{Diversity of Malware Types in Dataset}

As the security vendor does not provide the type of each sample, we use Kaspersky to determine the type of malicious samples and Detect it Easy (DIE) tool~\cite{DetectitEasy} to determine the type of benign samples. In Table~\ref{datasetinformation}, we present the results of our analysis. The results show that our dataset includes diverse malware and benign samples, making the detection model generalized to existing types of malware and benign samples. It is worth mentioning that there are 3\,160 samples that we were not able to obtain their type (indicated as ``Other'' in Table~\ref{datasetinformation}). These include 3\,145 samples that are classified as benign and 15 samples whose analysis report is not available.


\begin{table}[t]
\footnotesize 
\caption{Dataset Description}
\label{datasetinformation}
 \centering
\begin{tabular}{|l|l|r|l|r|}
\hline
                      & {\bf Types}                       & {\bf \#Samples} & {\bf Types}                       & {\bf \#Samples} \\ \hline
\multicolumn{1}{|l|}{\multirow{10}{*}{Malicious Samples}} & Trojan                     & 16\,154          & Packed                   & 50 \\ \cline{2-5} 
\multicolumn{1}{|l|}{}                            & Virus                     & 12\,144           & HackTool                   & 44  \\ \cline{2-5} 
\multicolumn{1}{|l|}{}                            & Worm                     & 4\,242            & Exploit                    & 30 \\ \cline{2-5} 
\multicolumn{1}{|l|}{}                            & Adware                        & 3\,179          & Rootkit                       & 14 \\ \cline{2-5} 
\multicolumn{1}{|l|}{}                            & Backdoor                       & 2\,661          & Porn                  & 10   \\ \cline{2-5} 
\multicolumn{1}{|l|}{}                            & Risktool                    & 1\,290            & RemoteAdmin                   & 7  \\ \cline{2-5} 
\multicolumn{1}{|l|}{}                            & Downloader                    & 549           & PswTool                   & 5  \\ \cline{2-5} 
\multicolumn{1}{|l|}{}                            & Hoax                    & 153           & Monitor                   & 3  \\ \cline{2-5} 
\multicolumn{1}{|l|}{}                            & DangerousObject                      & 150          & NetTool                &2   \\ \cline{2-5}
\multicolumn{1}{|l|}{}                            & WebToolBar  & 119             & Eicar               &1   \\ \cline{2-5} 
 
\multicolumn{1}{|l|}{}                            & Other               &3\,160    & &         \\\hline
\multicolumn{1}{|l|}{\multirow{2}{*}{Benign Samples}}     & DLL                        & 37\,527          & GUI                        & 4\,432  \\ \cline{2-5} 
\multicolumn{1}{|l|}{}                            & Console                    & 669             & Driver                     & 114   \\ \hline
\end{tabular}
\vspace{-5ex}
\end{table}

\subsubsection{Ground-truth Label Verification}

\begin{figure}[t]
    \centering
     \subfloat[Malicious Samples.]{
        \includegraphics[width=0.4\linewidth]{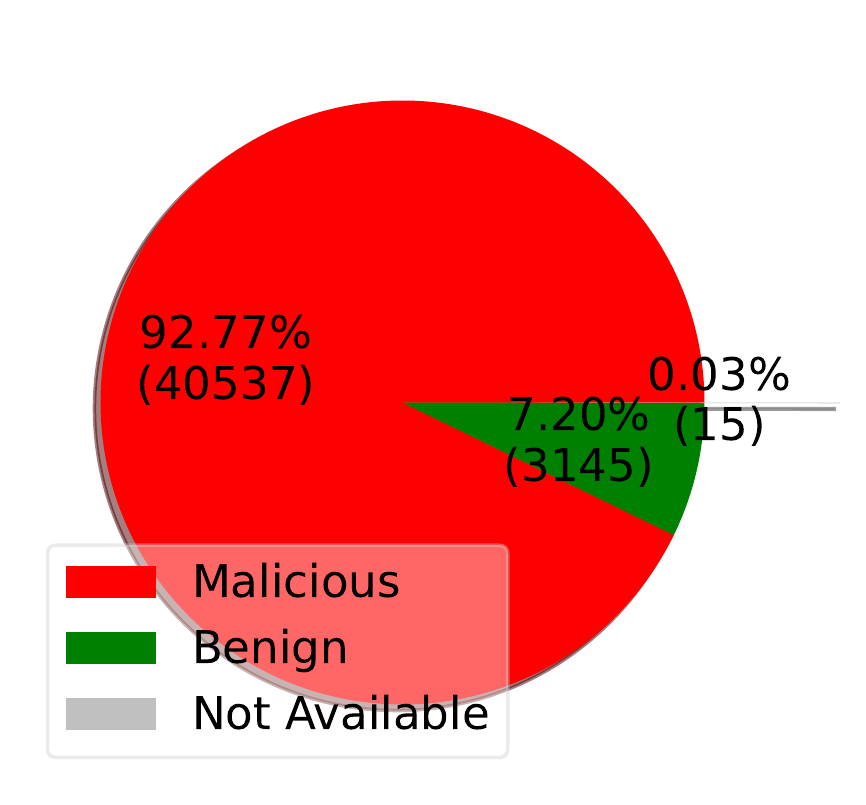}%
        \label{fig:label-matching-mal}
    }\hfill
    \subfloat[Benign Samples.]{
        \includegraphics[width=0.4\linewidth]{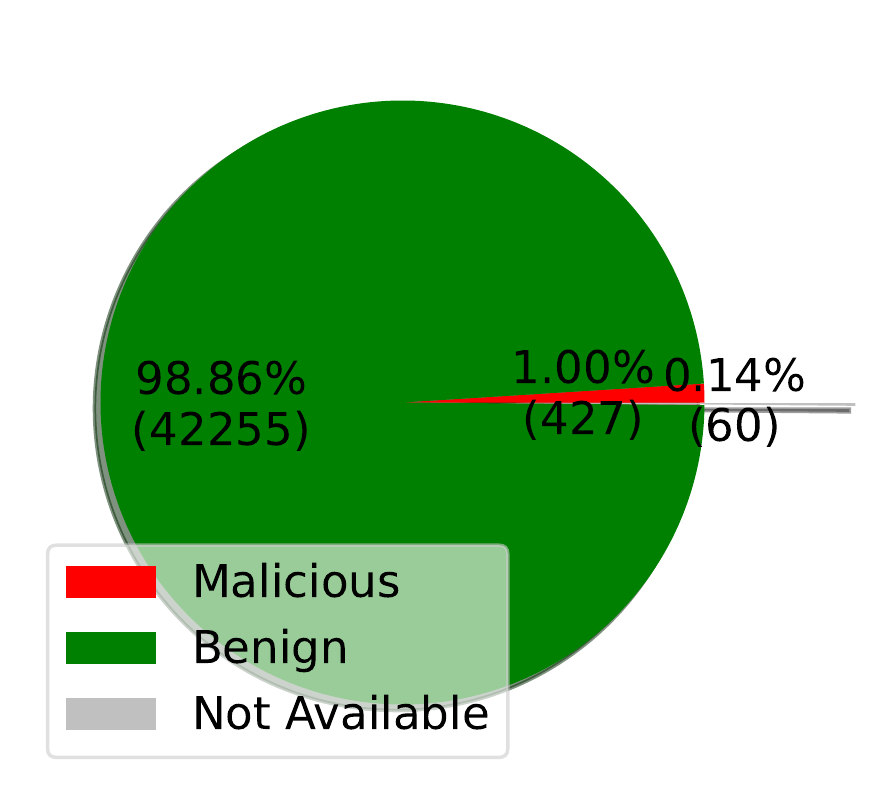}%
        \label{fig:label-matching-mal}
    }
    \caption{Matching of sample labels between Kaspersky and the anti-virus engine of the security vendor. Percentage of matching and the number of samples.}
    \label{fig:ground-truth}
    \vspace{-2ex}
\end{figure}

We use the labels provided by the security vendor to train the detection model in a supervised learning approach. To avoid the scenario that the detection model is biased toward the ground-truth labels of the security vendor, we use Kaspersky to verify the ground-truth labels provided by the security vendor. In Fig.~\ref{fig:ground-truth}, we present the matching between the ground-truth labels provided by the security vendor and the labels obtained from Kaspersky. We observe that the labels provided by the security vendor and Kaspersky agree for more than $92.77\%$ of malicious samples and $98.86\%$ of benign samples. There is $7.20\%$ of samples detected as malicious by the anti-virus engine of the security vendor but undetected by Kaspersky. There are also $1\%$ of samples considered benign by the anti-virus engine of the security vendor but detected as malicious by Kaspersky. Without loss of generality, we use the labels provided by the security vendor to train our detection models.

\subsection{Hyper-parameters Setting} 
\label{subsec:setupHyperparameters}

In this paper, we use stochastic gradient descent (SGD)\footnote{After experimenting with several optimizers, e.g., Adam, AdamW, RMSprop, SGD, and SGD with momentum, we observed that SGD with momentum performs best.} optimizer with a momentum of $0.9$ to train all the deep learning models for $100$ training epochs.
The initial learning rate is set as $0.02$ and reduced by $10$ times at the epoch $50$\textsuperscript{th} for EMBER, API-ARG, and aggregated feature vectors. For OPCODE features, we also use the initial learning rate of $0.02$ but reduce it by $10$ times at the epoch $30$\textsuperscript{th} and $80$\textsuperscript{th}. 
We also use a weight-decay of $0.001$ as a regularizer during the training of the models. Besides 1D-CNN models, we also train the XGBoost model, which is considered the state-of-the-art results in prior work~\cite{Dima_ACSAC2020}.
We use the default parameters of the open-source XGBoost Python-based package~\cite{xgboost}.

\subsection{Performance Metrics}
\label{subsec:performanceMetric}
We use accuracy, F1-score (F1-score=$\frac{2\text{TP}}{2\text{TP}+\text{FP}+\text{FN}}$) to measure the performance of different DL-based models. Moreover, for malware detection tasks, the false negative rate ($\text{FNR}=\frac{\text{FN}}{\text{FN}+\text{TP}}$) and false positive rate ($\text{FPR}=\frac{\text{FP}}{\text{FP}+\text{TN}}$) are also imperative metrics to compare the performance of detection models, where FN, FP, TN, TP are false negative, false positive, true negative, and true positive predicted samples, respectively. The smaller the values of FPR and FNR, the better the detection model.

\subsection{Performance of Teacher Model}
\label{subsec:ResultsTeacherModel}

\paragraph{\bf Experimental Results with Individual Feature Vectors}

Table~\ref{tab:separateFeatureResults} shows detection performance  on the test set of 1D-CNN models and XGBoost models\footnote{There is no variance of XGBoost's result because it does not include any stochastic in models; therefore yielding the same results regardless of running many times.} trained on individual feature vectors. We observe that for OPCODE and API-ARG features, the average performance of 1D-CNN models (over $5$ different trained models) is much higher than those of XGBoost models, which are state-of-the-art models used in~\cite{Anderson2018Ember,Dima_ACSAC2020}. For EMBER features, the average accuracy of 1D-CNN models is slightly lower than that of XGBoost models, but our best model (among the $5$ trained models) outperforms the XGBoost models. 

\begin{table}[t]
\small
    \caption{Performance (in percentage) on the test set of 1D-CNN (average of $5$ runs, and an ensemble model) and XGBoost}
    \label{tab:separateFeatureResults}
    \centering
    \begin{tabular}{ l@{\hspace{0.7em}} | l@{\hspace{0.4em}} | c@{\hspace{0.6em}}  c@{\hspace{0.6em}} c@{\hspace{0.5em}} c@{\hspace{0.5em}} }
    \hline
    \textbf{Feature}     & \textbf{Models} & \textbf{Accuracy} & \textbf{F1-score} & \textbf{FPR} & \textbf{FNR} \\
    \hline
    \multirow{2}{*}{\bf EMBER} &  1D-CNN & {97.62 $\pm$ 0.08} & {97.66 $\pm$ 0.08} & {2.46 $\pm$ 0.15} & 2.29 $\pm$ 0.12 \\
     & XGBoost & 97.66 & 97.70 & 2.55 & 2.13\\
    \hline
    \multirow{2}{*}{\bf OPCODE} & 1D-CNN & 95.91 $\pm$ 0.08 & 95.95 $\pm$ 0.08 & 3.78 $\pm$ 0.11 & 4.39 $\pm$ 0.11 \\
    & XGBoost & 95.54 &  95.59 & 4.16 & 4.74 \\
    \hline
    \multirow{2}{*}{\bf API-ARG} & 1D-CNN  & 96.64 $\pm$ 0.13 & 96.69 $\pm$0.13 & 3.29 $\pm$ 0.20 & 3.42 $\pm$ 0.23 \\
     & XGBoost & 96.48 & 96.55 & 4.01 & 3.03 \\
    \hline
    \end{tabular}
    \vspace{-2ex}
\end{table}

\paragraph{\bf Experimental Results with Aggregated Feature Vectors}

Among three individual feature vectors shown in Table~\ref{tab:separateFeatureResults}, the models trained on the EMBER feature vector outperform the models trained on OPCODE and API-ARG feature vectors by a significant margin. Therefore, we always use EMBER features in aggregated features. We conduct two ways of feature aggregation: 
\begin{itemize}
    \item \texttt{Agg2} concatenates EMBER and API-ARG;
 \item \texttt{Agg3} concatenates EMBER, OPCODE, and API-ARG.    
\end{itemize} 
 
With two feature aggregation methods described in Section~\ref{subsec:ArchitectureAgg}, we evaluate $4$ possible aggregated features as follows:
\begin{itemize}
    \item \texttt{Agg2-Org} concatenates original feature vectors of EMBER and API-ARG,
    \item \texttt{Agg2-Lat} combines latent representation vectors of EMBER and API-ARG,
    \item \texttt{Agg3-Org} combines original vectors of EMBER, OPCODE and API-ARG,
    \item \texttt{Agg3-Lat} concatenates latent representation vectors of EMBER, OPCODE and API-ARG.
\end{itemize}


Table~\ref{tab:ResultsAggModels} shows experimental results of 1D-CNN models and the XGBoost model. We can see that 1D-CNN models of \texttt{Agg2-Lat} and \texttt{Agg3-Lat} obtain a detection accuracy (both average and ensemble) higher than those of \texttt{Agg2-Org} and \texttt{Agg3-Org}, respectively. 
It is because \texttt{Agg2-Lat} and \texttt{Agg3-Lat} combine equally-dimensional latent representation vectors that learn important latent features from the original feature vectors. Specifically, in \texttt{Agg2-Org} and \texttt{Agg3-Org}, the original feature vector of API-ARG with high dimension significantly dominates other static original feature vectors. In contrast, \texttt{Agg2-Lat} and \texttt{Agg3-Lat} concatenate the latent representation vectors with the same dimension. The 1D-CNN model trained on \texttt{Agg2-Lat} feature vector obtains an average test accuracy of $97.95\%$ that is higher than the performance of the 1D-CNN model trained with \texttt{Agg2-Org}; and its ensemble model achieves the highest accuracy, F1-score, and the lowest FNR among all the evaluated models (i.e., $98.11\%$ for accuracy, $98.14\%$ for F1-score, $1.88\%$ for FPR and $1.89\%$ for FNR).  

We also observe that including more static features (i.e., concatenating both EMBER and OPCODE feature vectors) into the aggregated feature vector does not always help improve the performance of the models. Specifically, compared to \texttt{Agg2-Org}, \texttt{Agg3-Org} only improves accuracy and F1-score for 1D-CNN models but reduces performance for XGBoost models as shown in Table~\ref{tab:ResultsAggModels}.
Similarly, compared to \texttt{Agg2-Lat}, \texttt{Agg3-Lat} deteriorates in all performance metrics for both 1D-CNN and XGBoost models. This experiment shows that OPCODE features have negative effects on the aggregated features, especially when aggregating latent representation vectors with the same dimension. Thus, we need to carefully select individual feature vectors that help improve the performance of the models trained on the aggregated feature vector. Among all the methods, the ensemble of the models trained with \texttt{Agg2-Lat} obtains the best accuracy of $98.11\%$. We use this ensemble as the teacher model for transfer learning.

\begin{table}[t]
\small
    \caption{Performance (in percentage) of Teacher models trained with different aggregated feature vectors}
    \label{tab:ResultsAggModels}
    \centering
    \begin{tabular}{  l@{\hspace{0.6em}}|  c@{\hspace{0.6em}} | c@{\hspace{0.6em}} c@{\hspace{0.6em}}  c@{\hspace{0.6em}}  c@{\hspace{0.6em}} }
    \hline
     \textbf{Feature}    & \textbf{Models} & \textbf{Accuracy} & \textbf{F1-score} & \textbf{FPR} & \textbf{FNR} \\
    \hline
     \multirow{3}{*}{\textbf{Agg2-Org}}& 1D-CNN & 97.68 $\pm$ 0.05 & 97.72 $\pm$ 0.05 & 2.56 $\pm$ 0.15 & 2.08 $\pm$ 0.11 \\
      & Ensemble & 97.89 & 97.76 & 2.11 & 2.42  \\
                  & XGBoost & 97.99 & 98.02 & 2.15 & 1.87 \\
    \hline
     \multirow{2}{*}{\textbf{Agg2-Lat}} & 1D-CNN & {97.95 $\pm$ 0.04} & 97.98 $\pm$ 0.04  & 2.18 $\pm$ 0.11 & 1.92 $\pm$0.12  \\
      & Ensemble & \textbf{98.11} & \textbf{98.14} & 1.88 & \textbf{1.89} \\
     \hline
    \multirow{3}{*}{\textbf{Agg3-Org}} & 1D-CNN & 97.85 $\pm$ 0.03  &  97.88 $\pm$ 0.03 &  2.16 $\pm$ 0.08 & 2.15 $\pm$ 0.08  \\
     & Ensemble & 97.93 & 97.96 & 2.00 & 2.14 \\
                  & XGBoost & 97.96 & 97.99 & 2.21 & 1.87 \\
    \hline
    \multirow{2}{*}{\textbf{Agg3-Lat}} & 1D-CNN & 97.90 $\pm$ 0.06 & 97.93 $\pm$ 0.06 & 2.20 $\pm$ 0.08 & 2.00 $\pm$ 0.09 \\
    & Ensemble & 98.05 & 98.08 & \textbf{1.82} & 2.07\\
    \hline
    \end{tabular}
\end{table}

\subsection{Transfer Learning for EMBER Features}
\label{subsec:transferEmber}
In this section, we present experimental results of transfer learning from the teacher model trained with \texttt{Agg2-Lat} to a distilled student model trained with EMBER features. First, we fine-tune the scaling temperature hyper-parameter $\tau$ and the weighted scaling factor $\alpha$ for the loss functions in Eq.~\eqref{eq:lossesKL} and Eq.~\eqref{eq:lossesMSE}, then we evaluate the performance of the distilled student model.

\begin{figure}[t]
    \centering
    \subfloat[With different weighted parameter $\alpha$ in KD-KL loss and KD-MSE loss functions.]{
    \includegraphics[width=0.49\linewidth]{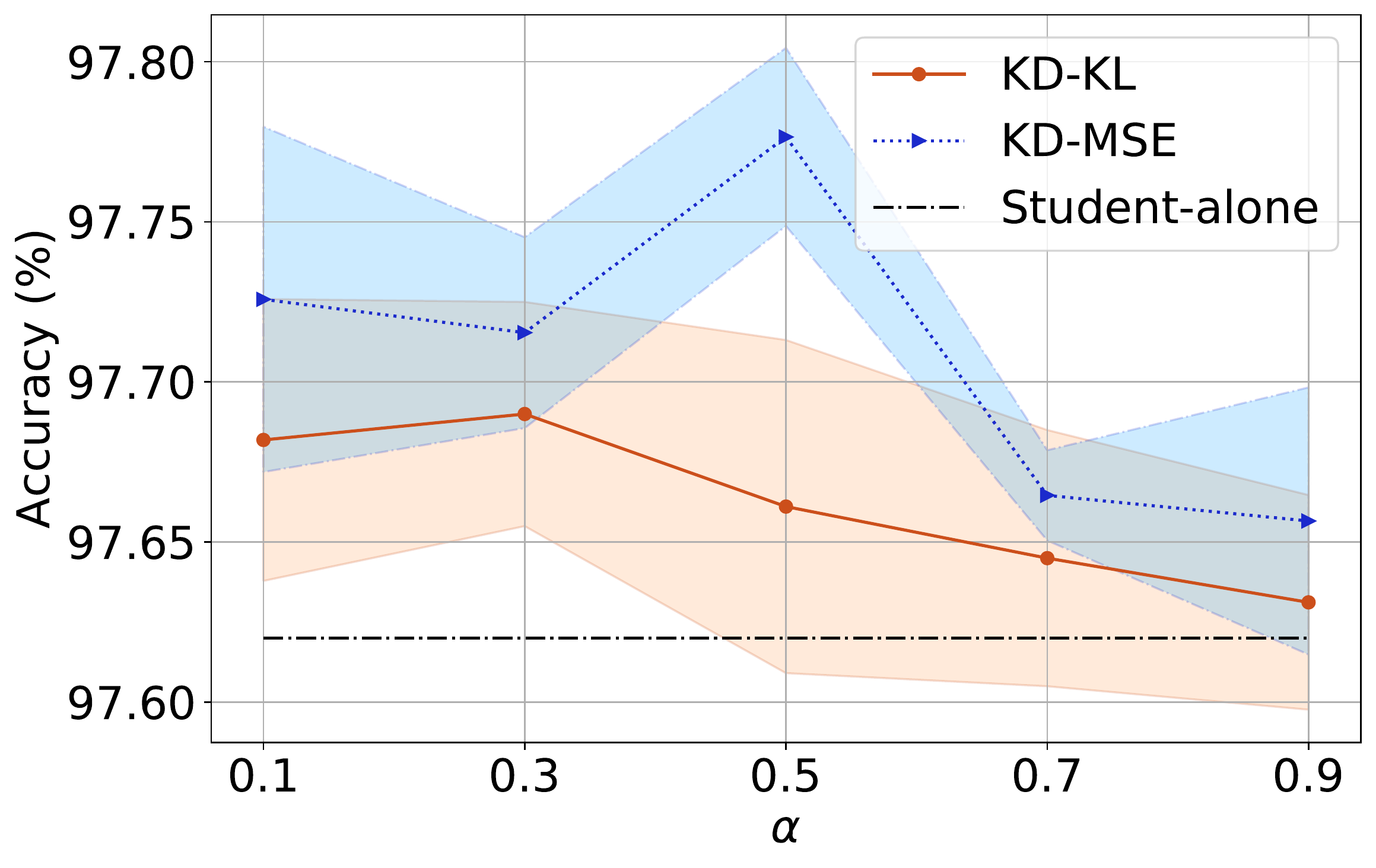}
    \label{fig:kdKL_MSE_TuneAlpha_Ember}}
    \subfloat[With different values of $\tau$ in KD-KL loss (with the best $\alpha=0.3$).]{
      \includegraphics[width=0.49\linewidth]{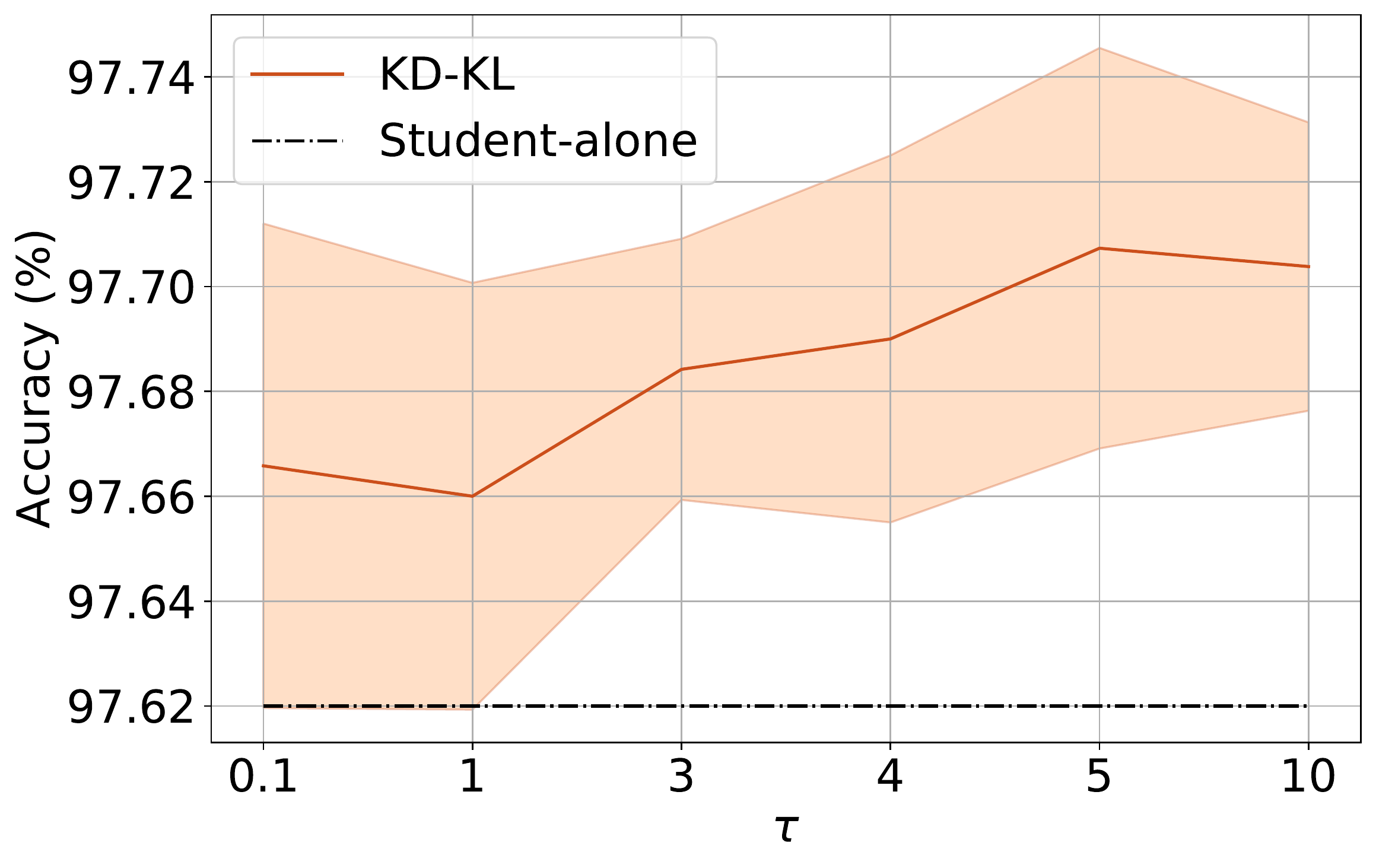} 
    \label{fig:kdKLTuneTauEmber}}
    \caption{Performance of Transfer Learning with EMBER Features.}
    \vspace{-2ex}
\end{figure}

\paragraph{\bf Tuning $\alpha$ for KD-KL loss and KD-MSE loss functions}
As suggested in~\cite{hinton2015distilling}, the weighted parameter $\alpha$ should be $0.1$, and the temperature $\tau \in \{3,4,5\}$. Nevertheless, we conduct a grid search to find the best $\alpha$ for our dataset. We transfer the knowledge from the teacher model to the distilled student models with EMBER features, with $\alpha \in \{0.1, 0.3, 0.5, 0.7, 0.9\}$ and a fixed temperature $\tau=4$. Fig.~\ref{fig:kdKL_MSE_TuneAlpha_Ember} shows the accuracy of these distilled students. We observe that the distilled models using the KD-MSE loss perform better than the distilled models using the KD-KL loss. This result aligns well with the results presented in~\cite{CompareKLMSE_IJCAI2021}. We also observe that the distilled student models with a small value of $\alpha$ perform better than those with a large value of $\alpha$ for both loss functions; it is because the distilled student models can learn more transferred knowledge from the ``soft'' labels predicted by the teacher models. Thus, we use the best value of $\alpha$ for KD-KL ($\alpha=0.3$) and KD-MSE ($\alpha=0.5$) for further evaluation. This result also aligns with the results presented in~\cite{hinton2015distilling}.

\paragraph{\bf Tuning scaling temperature hyper-parameter $\tau$ for KD-KL Loss}
With the selected value of $\alpha=0.3$ for KD-KL, we run experiments with different scaling temperatures $\tau \in\{0.1, 1, 3, 5 , 7, 10\}$. The mean and standard deviation of the accuracy of the distilled KD-KL student models for EMBER features (average over $5$ runs) are shown in Fig.~\ref{fig:kdKLTuneTauEmber}.
We can see that when increasing $\tau$'s value, the accuracy of the distilled student model slightly improves. This performance trend aligns with results from~\cite{CompareKLMSE_IJCAI2021}. The best accuracy is achieved with $\tau=5$ that aligns with the recommendation in~\cite{hinton2015distilling}.

In Table~\ref{tab:comparisonWithTransferLearningEmber}, we present a performance comparison between the student-alone model (i.e., the model trained with EMBER features without KD), and transferred KD models with KD-KL loss and KD-MSE loss (with their best hyper-parameters). We observe that the distilled student models outperform the student-alone model in all the metrics. The best-distilled student model (KD-MSE) obtains an accuracy of $97.81\%$, which is $0.11\%$ higher than that of the best student-alone model. Compared to the XGBoost model trained with OPCODE features (shown in Table~\ref{tab:separateFeatureResults}), this distilled student model improves the detection accuracy by $2.38\%$. The results also show that the performance of the distilled student models approximates that of the teacher models (i.e., $97.81\%$ vs. $98.11\%$) and is even better than the model trained with the aggregated feature vector (i.e., 1D-CNN with \texttt{Agg2-Org} achieves an accuracy of $97.68\%$ shown in Table~\ref{tab:ResultsAggModels}). 

\begin{table}[t]
\small
    \caption{Performance (in percentage) of student models transferred with different loss functions for EMBER features}
    \label{tab:comparisonWithTransferLearningEmber}
    \centering
    \begin{tabular}{ l@{\hspace{0.4em}} | c@{\hspace{0.4em}} c@{\hspace{0.4em}} c@{\hspace{0.4em}} c@{\hspace{0.4em}} }
    \hline
    \textbf{Methods} &  \textbf{Accuracy} & \textbf{F1-score} & \textbf{FPR} & \textbf{FNR}\\
    \hline
    \textbf{Student-alone} & 97.62 $\pm$ 0.08 & 97.66 $\pm$0.12 & 2.45 $\pm$ 0.15  & 2.29 $\pm$0.12 \\
    \hline
    \textbf{KD-KL} & 97.71 $\pm$ 0.04 & 97.74$\pm$ 0.04 & 2.38 $\pm$0.09 & 2.21 $\pm$ 0.14 \\ %
    \textbf{KD-MSE} & \textbf{97.78 $\pm$ 0.03} & \textbf{97.81 $\pm$ 0.03} & \textbf{2.26 $\pm$ 0.07} & \textbf{2.19 $\pm$ 0.08} \\ %
    \hline
    \end{tabular}
    \vspace{-2ex}
\end{table}

When aggregating latent representations of EMBER features and dynamic API-ARG features, \texttt{Agg2-Lat} can improve $0.33\%$ of accuracy compared to the EMBER student-alone ($98.11\%$ in Table~\ref{tab:ResultsAggModels} vs. $97.62\%$ in Table~\ref{tab:comparisonWithTransferLearningEmber}). But this incurs a significant delay in inference due to dynamic analysis and results in a large model (discussed further in Section~\ref{subsec:E2Edelay}).
When using knowledge distillation to train a distilled KD-MSE student model, we obtain an improvement of $0.16\%$ of accuracy, which is about $50\%$ of the improvement of the model trained with \texttt{Agg2-Lat} but \textit{without any additional cost or delay due to dynamic analysis}. The model also keeps unchanged in terms of size and inference time. It is to be noted that even though the absolute number of $0.16\%$ of improvement is quite small, the standard deviation of the result is very small ($0.03\%$), which shows reliable and consistent improvement.
 
\subsection{Transfer Learning with OPCODE Features}
\label{subsec:resultOPCODE}

We also study KD from the teacher model to a 1D-CNN student model trained with OPCODE features.

\begin{figure}[t]
    \centering
    \subfloat[With different values of hyper-parameter $\alpha$ in KL-loss (with $\tau=4.0$) and MSE-loss.]{
    \includegraphics[width=0.49\linewidth]{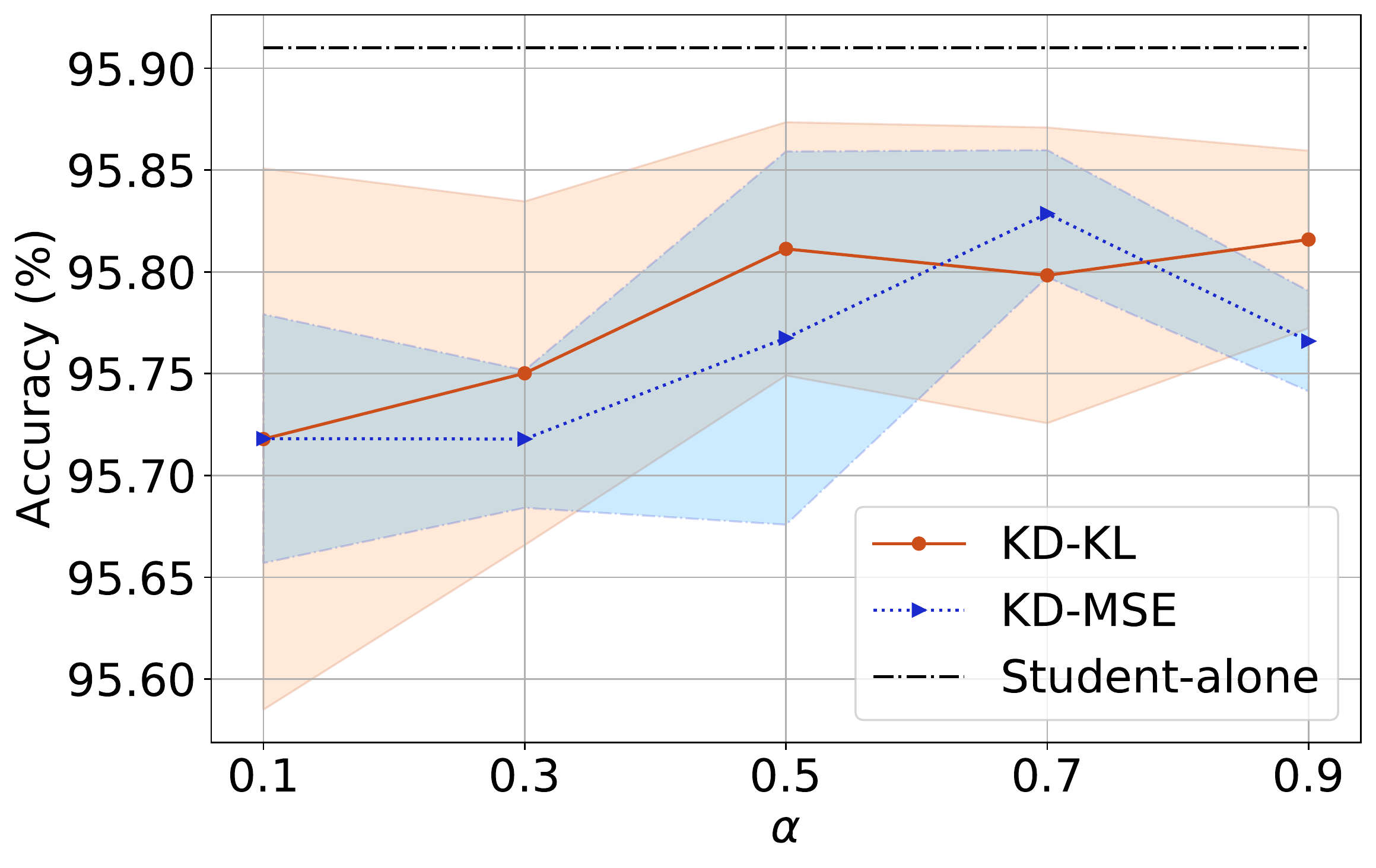}
    \label{fig:kdKL_kdMSE_TuneAlphaOPCODE}}
    \subfloat[With different hyper-parameter $\tau$ in KL-loss (with $\alpha=0.5$).]{
      \includegraphics[width=0.49\linewidth]{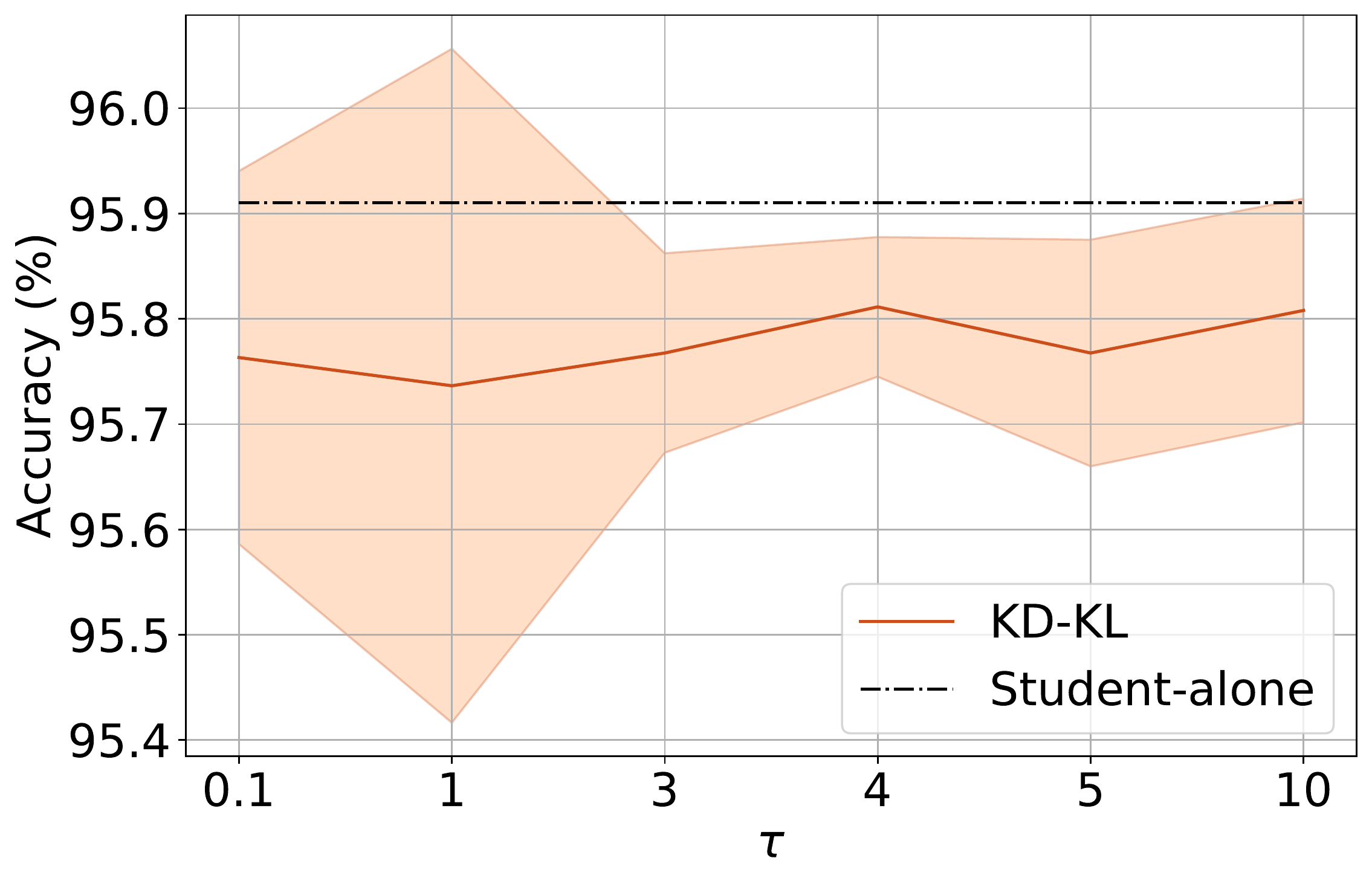} 
    \label{fig:kdKL_Tune_T_OPCODE}}
    \caption{Performance of Transfer Learning with OPCODE Features.}
    \vspace{-2.5ex}
\end{figure}

\paragraph{\bf Tuning $\alpha$ for KD-KL loss and KD-MSE loss functions}
We conduct a similar fine-tuning of the $\alpha$ hyper-parameter for KD-KL loss and KD-MSE loss by using the same teacher as used for EMBER features (i.e., \texttt{Agg2-Lat}). The results are shown in Fig.~\ref{fig:kdKL_kdMSE_TuneAlphaOPCODE}. We observe an interesting result that regardless of $\alpha$ for both loss functions, all the distilled student models trained with OPCODE features obtain a lower accuracy than the OPCODE student-alone model even including the upper bound of the shaded areas.

\paragraph{\bf Tuning scaling temperature hyper-parameter $\tau$ for KD-KL Loss}
We fine-tune temperature hyper-parameter $\tau$ with the best $\alpha=0.5$ obtained in Fig.~\ref{fig:kdKL_kdMSE_TuneAlphaOPCODE}. As shown in Fig.~\ref{fig:kdKL_Tune_T_OPCODE}, the average accuracy of the distilled student models trained with OPCODE features and all the evaluated values of $\tau$ is lower than that of the student-alone model. We conjecture that the KD technique can only transfer knowledge from a teacher to a student model trained with a feature vector that has a positive contribution to the teacher model (e.g., EMBER); while it fails to do so if the feature vector used to train the student model has a negative contribution to the teacher model (e.g., OPCODE) when aggregation. We remind the reader about the results presented in Section~\ref{subsec:ResultsTeacherModel}, \texttt{Agg3-Lat} (with OPCODE) performs worse than \texttt{Agg2-Lat} (without OPCODE). It is also to be noted that the student-alone model trained with OPCODE features without KD has the lowest performance compared to the models trained with EMBER features or API-ARG features. This aligns well with the analogy that a student who is not strong at learning will not be able to acquire knowledge from the teacher even though the teacher is strong.

\begin{table}[t]
\small
    \caption{Comparison between student-alone model and distilled models trained with OPCODE features}
    \label{tab:comparisonWithTransferLearningOpcode}
    \centering
    \begin{tabular}{ l@{\hspace{0.6em}} | r@{\hspace{0.6em}}  r@{\hspace{0.6em}} r@{\hspace{0.6em}} r@{\hspace{0.6em}} }
    \hline
    \textbf{Methods} &  \textbf{Accuracy} & \textbf{F1-score} & \textbf{FPR} & \textbf{FNR} \\
    \hline
    \textbf{Student-alone} & \textbf{95.91 $\pm$ 0.08} & \textbf{95.95$\pm$ 0.08}  & \textbf{3.77 $\pm$ 0.11} & 4.40 $\pm$ 0.11 \\
    \hline
    \textbf{KD-KL-2F} & 95.81 $\pm$ 0.11 & 95.86$\pm$0.11  & 4.06$\pm$0.15 & \textbf{4.32$\pm$0.14} \\ 
    \textbf{KD-MSE-2F} & 95.83 $\pm$ 0.03 &  95.87$\pm$0.03 & 3.93$\pm$ 0.09 & 4.40$\pm$0.07\\ 
    \hline
    \textbf{KD-KL-3F} & 95.75 $\pm$ 0.04 & 95.80$\pm$0.04  & 4.07$\pm$ 0.17 & 4.43$\pm$0.16 \\ 
    \textbf{KD-MSE-3F} & 95.78 $\pm$ 0.09 & 95.83 $\pm$0.09  & 3.96$\pm$0.14 & 4.48$\pm$0.11 \\  
    \hline
    \end{tabular}
    \vspace{-2.5ex}
\end{table}

Since the results from Fig.~\ref{fig:kdKL_kdMSE_TuneAlphaOPCODE} and Fig.~\ref{fig:kdKL_Tune_T_OPCODE} were obtained by transferring the knowledge from the teacher trained without OPCODE features to a student using OPCODE features, we conduct further experiments with transfer learning from a teacher with OPCODE features included in the aggregated feature vector (i.e., \texttt{Agg3-Lat}) to a student model trained with OPCODE features. We use the best hyper-parameter obtained above ($\alpha=0.5, \tau=4$ for KD-KL, and $\alpha=0.7$ for KD-MSE) to conduct these experiments. The last two rows in Table~\ref{tab:comparisonWithTransferLearningOpcode} show the performance (in percentage) of the distilled student models trained with OPCODE features (i.e., KD-KL-3F and KD-MSE-3F) with knowledge transferred from softened labels obtained from the teacher model trained with \texttt{Agg3-Lat} feature vector. We can see that the performance of KD-KL-3F and KD-MSE-3F is even worse than that of KD-KL-2F and KD-MSE-2F. This is because the teacher model trained with \texttt{Agg3-Lat} feature vector performs poorer than the teacher model trained with \texttt{Agg2-Lat} feature vector. This aligns with another analogy that a poor teacher cannot transfer his/her knowledge to students and could deteriorate the performance of the students if the knowledge is not effective. 
\subsection{End-to-End Detection Delay}
\label{subsec:E2Edelay}

In Table~\ref{tab:breakdownE2EDelayDetection}, we report a breakdown of end-to-end (E2E) detection delay that consists of three components: (i) malware analysis delay, (ii) feature extraction (\textit{Feat-Extr.}) delay, and (iii) inference delay for different detection methods. We can see that (1) our proposed method---KD-MSE and EMBER method have the least E2E delay for detecting malware (i.e., $194.9$\,ms); and (2) detection methods involving dynamic analysis (i.e., API-ARG, and \texttt{Agg2-Lat}) have significantly long E2E delay compared to that of other methods (OPCODE, EMBER, and KD-MSE), at least $7\times{}$ longer. 

\begin{table}[t]
    \caption{Breakdown of end-to-end detection delay (in ms)}
    \label{tab:breakdownE2EDelayDetection}
    \centering
    \begin{tabular}{ l@{\hspace{0.6em}} | r@{\hspace{0.6em}}  |r@{\hspace{0.6em}} | r@{\hspace{0.6em}} || r@{\hspace{0.6em}} }
    \hline
    \textbf{Methods} & \textbf{Analysis}  &  \textbf{Feat-Extr.}  & \textbf{Inference}  & \textbf{E2E Delay} \\
    \hline
    \textbf{OPCODE}&  12\,243.7 & 62.0 & 0.4 & 12\,306.1 \\ \hline
    \textbf{API-ARG}  & 69\,120.0$^{(a)}$ & 720.0  & 1.2 & 69\,841.2  \\
    \hline
    \textbf{Agg2-Lat}&  69\,120.0  & 914.7 & 11.9$^{(b)}$ & 70\,046.6 \\
    \hline
    \textbf{EMBER} &  \multicolumn{2}{c|}{194.7$^{(c)}$} & 0.2 & \textbf{194.9} \\ \hline
    \textbf{KD-MSE} &  \multicolumn{2}{c|}{194.7$^{(d)}$} & 0.2 & \textbf{194.9} \\
    \hline
    \multicolumn{5}{p{10cm}}{\footnotesize{$^{(a)}$The delay of dynamic analysis of malicious samples is usually much higher than those of benign samples, e.g., the average delay to complete analysis of a malicious sample is $120\,960$ ms, while this figure for a benign sample is only $60\,480$ ms.}} \\
    \multicolumn{5}{p{10cm}}{\footnotesize{$^{(b)}$Inference delay of the ensemble model from $11$ models.}}\\
    \multicolumn{5}{p{10cm}}{\footnotesize{$^{(c)}$For EMBER features, we measure analysis delay and feature-extraction delay jointly in a single program.}}\\
    \multicolumn{5}{p{10cm}}{\footnotesize{$^{(d)}$Analysis delay for KD-MSE is similar to EMBER as it uses EMBER feature for detection.}}
    \end{tabular}
\end{table}

Looking into each delay component of API-ARG and \texttt{Agg2-Lat} methods, we see the \textit{Feat-Extr.} and \textit{Inference} delays are minuscule compared to \textit{Analysis} delay\footnote{Analysis delays are calculated as the average time of executing samples in the test set (i.e., $17\,342$ samples) using dynamic or static analysis.} (which consumes above 98\% of E2E delay). Thus, removing dynamic \textit{Analysis} delay will significantly speed up malware detection time in a real-world deployment. This explains why static analysis is widely used in many commercial-off-the-shelf antivirus products. The experimental results also show that among the static analysis-based methods, OPCODE needs a much longer time for inference compared to EMBER and KD-MSE in both feature extraction and inference. In summary, our KD-MSE approach achieves a detection as fast as the fastest static method (i.e., EMBER), while approximating the performance of the \texttt{Agg2-Lat} method that uses both dynamic and static features.

It is worth mentioning that reducing the inference/detection overhead is crucial for a detection model as delaying the detection affects the usability of the system and raises security issues: malware samples may compromise the host before being detected. Nevertheless, we note that the training time of the models (e.g., teacher model and student model) is a one-time cost and it could be done offline with sufficient computing resources. This reflects practical scenarios where security companies nowadays deploy large-scale computing systems in their premise for pre-training and updating detection models, which are in turn deployed on commodity computers of their end-users.

\section{Conclusion}
\label{sec:conclusion}
In this paper, we developed various 1D-CNN models for malware detection using static and dynamic features. We addressed a limitation of the feature aggregation that naively concatenates original static and dynamic feature vectors by using a 1D-CNN model to learn the latent representation of each original feature vector with a fixed size before aggregation. We tackled the dilemma between the benefits of dynamic analysis and its massive delay in the deployment phase by developing a \textit{knowledge distillation} technique to \textit{transfer} rich knowledge from a big teacher model (trained with aggregated features) to a small student model (trained with static features only). We carried out extensive experiments with a dataset of $86\,709$ samples to evaluate the performance and efficiency of our proposed technique. The experimental results show that the proposed technique outperforms existing methods with an improvement of up to $2.38\%$ in terms of detection accuracy. The proposed technique also significantly reduces the end-to-end delay for sample prediction from $70\,046.6$ms to $194$ms. In other words, the proposed technique enables fast and efficient malware detection. Our work also found that transfer learning is not always successful if the student feature vector has negative effects on the performance of the teacher when aggregated with the teacher feature vector. 

\bibliographystyle{splncs04}
\bibliography{references}


\appendix
\section{Ablation Studies}
\label{sec:ablationStudy}

\subsection{Transfer Learning for API-ARG Features}
\label{sec:Transfer_APIARG}
Even though transferring knowledge to a student model trained with dynamic features is not a focus of our work, we evaluate the performance of the distilled student models trained with dynamic API-ARG features in this section for completeness. Since the training time for a student model trained with API-ARG features is much longer than that for the models trained with EMBER features or OPCODE features, we only evaluate one set of the recommended hyper-parameters from~\cite{hinton2015distilling}: $\alpha=0.1, \tau=5$ for KD-KL, and $\alpha=0.1$ for KD-MSE. As shown in Table~\ref{tab:resultKD_APIARG}, the distilled student models for API-ARG features perform better than the student-alone model. The performance trends also follow our conjecture that knowledge distillation helps transfer knowledge to a distilled student model with a dynamic API-ARG feature vector that \textit{positively contributes} to the teacher model (i.e., \texttt{Agg2-Lat}). We also observe that the KD-KL student model obtains better performance compared to the KD-MSE student model, which is different from the results of the case where the student model is trained with EMBER features. This explains why we explored two loss functions during transfer learning for completeness, even though in~\cite{CompareKLMSE_IJCAI2021} the authors stated that the KL-loss function is more suitable for a noisy label case, which is not our case since the labels of the malware samples are stabilized and provided by the security vendor.

\begin{table}[t]
\small 
    \caption{Performance (in percentage) of API-ARG student models transferred with different loss functions}
    \label{tab:resultKD_APIARG}
    \centering
    \begin{tabular}{ l@{\hspace{0.6em}} | c@{\hspace{0.6em}} c@{\hspace{0.6em}} c@{\hspace{0.6em}} c@{\hspace{0.6em}} }
    \hline
    \textbf{Methods} &  \textbf{Accuracy} & \textbf{F1-score} & \textbf{FPR} & \textbf{FNR}\\
    \hline
    \textbf{Student-alone} & 96.64 $\pm$ 0.13 & 96.69 $\pm$0.13 & 3.29 $\pm$ 0.20  & 3.42 $\pm$0.23 \\
    \hline
    \textbf{KD-KL} & \textbf{96.77 $\pm$ 0.02} & \textbf{96.81$\pm$ 0.02} & \textbf{3.19 $\pm$ 0.07} & 3.26 $\pm$ 0.06 \\ %
    \textbf{KD-MSE} & 96.73 $\pm$ 0.08 & 96.77 $\pm$ 0.08 & 3.29 $\pm$ 0.13 & \textbf{3.24 $\pm$ 0.08} \\ %
    \hline
    \end{tabular}
\end{table}

\subsection{Experiments with Different Neural Network Architectures}
\label{subsec:ablationNetwork}

\begin{figure}[t]
    \centering
     \subfloat[ResNet1D block with kernel size $K=3$]{
        \includegraphics[width=0.3\linewidth]{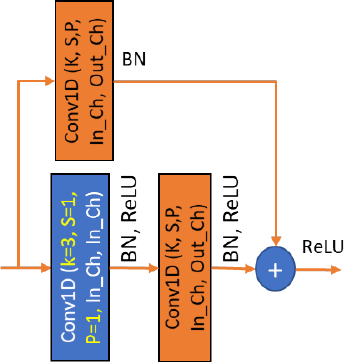}%
        \label{fig:ResNetBasicBlock}
    }
    \subfloat[ResNeXt1D \& Inverted ResNeXt1D blocks~\cite{xie2017ResNeXt}]{
        \includegraphics[width=0.36\linewidth]{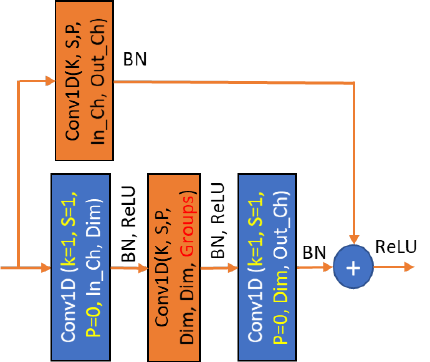}%
        \label{fig:ResNextBlock}
    }
    \subfloat[ConvNeXt block~\cite{liu2022convnet}]{
        \includegraphics[width=0.34\linewidth]{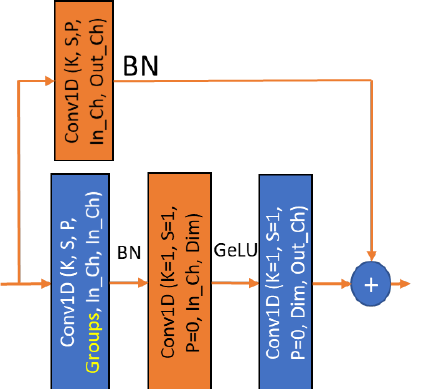}%
        \label{fig:ConvNextBlock}
    }
    \caption{Different block architectures for 1D-CNN models.}
    \label{fig:networkArchitectureDiffBlock}
    \vspace{-2.5ex}
\end{figure}

Besides ResNet1D used for all the experiments presented above, we adopt recent advanced neural network architectures of ResNet in computer vision such as ResNeXt~\cite{xie2017ResNeXt}, inverted ResNeXt~\cite{xie2017ResNeXt}, and recently ConvNeXt~\cite{liu2022convnet}. We adjust our ResNet1D basic block and implement its variants, namely ResNeXt1D, Inverted ResNeXt1D, and ConvNeXt1D basic blocks accordingly, presented in Fig.~\ref{fig:networkArchitectureDiffBlock}. We reuse the high-level neural network architectures as defined for ResNet1D in Table~\ref{tab:detailedNetworkArchitectEmber}, Table~\ref{tab:detailedNetworkArchitectOpcode}, and Table~\ref{tab:detailedNetworkArchitectAPIARG} for each individual feature vector (i.e., EMBER, OPCODE, and API-ARG). 

The average performance over $5$ runs of ResNet1D (with kernel size $K=3$ and $K=1$) and other variants of the student-alone models trained with EMBER, OPCODE features, and the teacher model trained with \texttt{Agg2-Lat} feature vector are presented in  Table~\ref{tab:differentResidualBlocks}. 
We observe that ResNet1D with kernel size $K=3$ obtains the highest accuracy and small size for all the models trained with EMBER, OPCODE, and \texttt{Agg2-Lat} feature vectors.
This explains why we chose ResNet1D with kernel size $K=3$ as our basic block for all the experiments presented above.

\begin{table}[t]
    \caption{Accuracy (in percentage) and size (in MB) of models with different types of basic blocks (see Fig.~\ref{fig:networkArchitectureDiffBlock}) for EMBER, OPCODE, and aggregated feature vectors}
    \label{tab:differentResidualBlocks}
    
    \centering
    \begin{tabular}{ l@{\hspace{0.6em}}  l@{\hspace{0.6em}} | c@{\hspace{0.6em}}  c@{\hspace{0.6em}} c@{\hspace{0.6em}}  }
    \hline
    {\bf Models} & & \textbf{EMBER} & \textbf{OPCODE} & \textbf{Agg2-Lat}\\
    \hline
    \textbf{ResNet1D} & Acc. & \textbf{97.62 $\pm$ 0.08} & \textbf{95.91 $\pm$ 0.08} & \textbf{97.95 $\pm$ 0.04} \\
    \textbf{($K=3$)}& Size & 7.51 & 16.67 & 271.27 \\
    \hline
    \textbf{ResNet1D} & Acc. & 97.61 $\pm$ 0.06 &  95.79 $\pm$ 0.08 & 97.91 $\pm$ 0.03 \\
    \textbf{($K=1$)} & Size & 7.11  & \textbf{16.27} & \textbf{269.98} \\
    \hline
    \multirow{2}{*}{\textbf{ResNeXt1D}} & Acc. & 97.57 $\pm$ 0.04 & 95.79 $\pm$ 0.11 & 97.88 $\pm$ 0.07 \\
    & Size & \textbf{5.44} & 28.84 & 405.03 \\ 
    \hline
    \textbf{Inverted}  & Acc. & 97.56 $\pm$ 0.06 & 92.68 $\pm$ 6.9 &  97.80$\pm$0.06 \\
    \textbf{ResNeXt1D} & Size  & 6.40 & 37.47 & 611.37 \\ 
    \hline
    \multirow{2}{*}{\textbf{ConvNeXt1D}} & Acc. & 97.08 $\pm$ 0.02 & 95.23 $\pm$ 0.05 & 97.40 $\pm$ 0.06 \\
    &  Size & 6.40 & 21.66 & 230.55 \\ 
    \hline
    \end{tabular}
\end{table}

\section{Implementation of Neural Network Architecture for Individual Feature Vectors}
\label{sec:AppArchtiectureIndividualFeatures}

Detailed implementation of neural network architectures for the EMBER feature vector, OPCODE feature vector and API-ARG feature vector are shown in Table~\ref{tab:detailedNetworkArchitectEmber}, Table~\ref{tab:detailedNetworkArchitectOpcode}, and Table~\ref{tab:detailedNetworkArchitectAPIARG}, respectively.

\begin{table}[t]
    \caption{Detailed network architecture for EMBER feature vector with $2\,381$ dimensions}
    \label{tab:detailedNetworkArchitectEmber}
    \centering
    \begin{tabular}{ l@{\hspace{0.6em}}| c@{\hspace{0.6em}} | c@{\hspace{0.6em}} |  c@{\hspace{0.6em}} |  c@{\hspace{0.6em}} }
    \hline
    \textbf{Block} & \textbf{Kernel size} & \textbf{Stride}   & \textbf{Padding} & \textbf{Out channels} \\ 
    \hline
    conv1 & 7 & 4 & 1& 24 \\
    conv2 & 7 & 5 & 1 & 48 \\
    conv3 & 7 & 4 & 0 & 96 \\
    conv4 & 7 & 4 & 1 & 192 \\
    conv5 & 7 & 1 & 0 & 384 \\
    \hline
    FC1 & \multicolumn{4}{c}{384 x 128} \\
    FC2 & \multicolumn{4}{c}{128 x 2 } \\
    \hline
    \end{tabular}
\end{table}

\begin{table}[t]
    \caption{Detailed network architecture for OPCODE feature vector with $33\,338$ dimensions}
    \label{tab:detailedNetworkArchitectOpcode}
    \centering
    \begin{tabular}{ l@{\hspace{0.6em}}| c@{\hspace{0.6em}} | c@{\hspace{0.6em}} |  c@{\hspace{0.6em}} |  c@{\hspace{0.6em}} }
    \hline
    \textbf{Block} & \textbf{Kernel size} & \textbf{Stride} & \textbf{Padding} & \textbf{Out channels} \\ 
    \hline
    conv1 & 10 & 5 & 1 & 16 \\
    conv2 & 12 & 5 & 0 & 24 \\
    conv3 & 12 & 5 & 0 & 48 \\
    conv4 & 10 & 5 & 0 & 96 \\
    conv5 & 12 & 5 & 0 & 192 \\
    conv6 & 9 & 1 & 0 & 384 \\
    \hline
    FC1 & \multicolumn{4}{c}{384 x 128} \\
    FC2 & \multicolumn{4}{c}{128 x 2 } \\
    \hline
    \end{tabular}
\end{table}

\begin{table}[t]
    \caption{Detailed network architecture for API-ARG feature vector with $1\,048\,576$ dimensions}
    \label{tab:detailedNetworkArchitectAPIARG}
    \centering
    \begin{tabular}{ l@{\hspace{0.6em}}| c@{\hspace{0.6em}} | c@{\hspace{0.6em}} |  c@{\hspace{0.6em}} |  c@{\hspace{0.6em}} }
    \hline
    \textbf{Block} & \textbf{Kernel size} & \textbf{Stride} & \textbf{Padding}  & \textbf{Out channels} \\ 
    \hline
    conv1 & 11 & 5 & 0 & 16 \\
    conv2 & 11 & 5 & 1 & 24 \\
    conv3 & 12 & 7 & 0 & 48 \\
    conv4 & 11 & 5 & 0 & 72 \\
    conv5 & 11 & 7 & 2 & 96 \\
    conv6 & 11 & 8 & 0 & 192 \\
    conv7 & 11 & 5 & 0 & 240 \\
    conv8 & 3 & 1 & 0 & 384 \\
    \hline
    FC1 & \multicolumn{4}{c}{384 x 128} \\
    FC2 & \multicolumn{4}{c}{128 x 2 } \\
    \hline
    \end{tabular}
\end{table}

\section{Neural Network Architecture of Aggregated Original Features}
\label{sec:ArchitectureAggRawFeatures}

In Table~\ref{tab:Agg2RawNet} and Table~\ref{tab:Agg3RawNet}, we present the architectures of the 1D-CNN models for aggregated feature vectors from $2$ original feature vectors (\texttt{Agg2-Org}---EMBER + API-ARG) and $3$ original feature vectors (\texttt{Agg3-Org}---EMBER + OPCODE + API-ARG), which are developed and evaluated in our work.

\begin{table}[t]
    \centering
    \caption{Detailed network architectures for \texttt{Agg2-Org} feature vector ($2\,381 + 1\,048\,576 = 1\,050\,957$ dimensions)}
    \label{tab:Agg2RawNet}
    \begin{tabular}{ l@{\hspace{0.5em}}| c@{\hspace{0.5em}} | c@{\hspace{0.5em}} |  c@{\hspace{0.5em}} |  c@{\hspace{0.5em}} }
    \hline
    \textbf{Block} & \textbf{Kernel size} & \textbf{Stride} & \textbf{Padding}  & \textbf{Out channels} \\ 
    \hline
    conv1 & 11 & 6 & 1 & 16 \\
    conv2 & 16 & 8 & 1 & 24 \\
    conv3 & 16 & 8 & 1 & 48 \\
    conv4 & 12 & 4 & 0 & 96 \\
    conv5 & 12 & 5 & 0 & 192 \\
    conv6 & 11 & 4 & 0 & 192 \\
    conv7 & 8 & 4 & 0 & 384 \\
    conv8 & 7 & 1 & 0 & 384 \\
    \hline
    FC1 & \multicolumn{4}{c}{384 x 128} \\
    FC2 & \multicolumn{4}{c}{128 x 2 } \\
    \hline
    \end{tabular}
\end{table}

\begin{table}[t]
    \centering
    \caption{Detailed network architectures for \texttt{Agg3-Org} feature vector ($2\,381+33\,338+1\,048\,576=1\,084\,295$ dimensions)}
    \label{tab:Agg3RawNet}
    \begin{tabular}{ l@{\hspace{0.6em}}| c@{\hspace{0.6em}} | c@{\hspace{0.6em}} |  c@{\hspace{0.6em}} |  c@{\hspace{0.6em}}}
    \hline
    \textbf{Block} & \textbf{Kernel size} & \textbf{Stride} & \textbf{Padding}  & \textbf{Out channels} \\ 
    \hline
    conv1 & 11 & 6 & 0 & 16 \\
    conv2 & 15 & 8 & 2 & 24 \\
    conv3 & 15 & 8 & 1 & 48 \\
    conv4 & 11 & 4 & 0 & 96 \\
    conv5 & 11 & 5 & 1 & 192 \\
    conv6 & 7 & 5 & 1 & 192 \\
    conv7 & 8 & 4 & 0 & 384 \\
    conv8 & 6 & 1 & 0 & 384 \\
    \hline
    FC1 & \multicolumn{4}{c}{384 x 128} \\
    FC2 & \multicolumn{4}{c}{128 x 2 } \\
    \hline
    \end{tabular}
\end{table}

\section{Speeding up Dynamic Analysis with Distributed Cuckoo Infrastructure}
\label{appendix:dist-cuckoo}
As we discussed earlier, it is a time-consuming task of dynamic analysis to produce analysis reports for malware samples. It depends on the user-defined parameter of the longest time that a sample is analyzed in the Cuckoo sandbox but the default setting is two minutes. Based on our daily experiments, using a single Cuckoo sandbox, we were able to analyze $8000$ samples per week. We could use multiple sandboxes and manually submit malware samples to speed up the analysis. However, this is quite laborious as the virtual machines hosting Cuckoo sandboxes frequently crash, thus requiring close monitoring for fixing occurring issues. 

In this work, we developed a parallel dynamic analysis infrastructure using the preliminary version of distributed Cuckoo~\cite{DistCuckoo}. We enriched the infrastructure by adding more automation for fault tolerance, which includes 
\begin{itemize}
\item A re-submission mechanism to resubmit the samples that have not been successfully analyzed. The number of resubmissions is a user-predefined parameter (e.g., three times).   
\item A monitoring mechanism to check the status of virtual machines whether they are in normal working conditions or crashed. We implement an active monitoring technique that periodically sends monitoring requests to Oracle VirtualBox~\cite{VirtualBox} to check virtual machine status (Cuckoo uses Oracle VirtualBox to host virtual machines and sandboxes). 
\item A virtual machine (VM) instantiation  mechanism to instantiate new VMs to replace the crashed ones. This process is automatically invoked when the monitoring system detects a crashed VM. This allows us to maximize the utilization of computing resources (i.e., the available VMs hosted on physical servers). 
\end{itemize}

With the developed infrastructure and the available resources in our lab, we could run up to $12$ Cuckoo sandboxes at the same time. The implemented fault-tolerance mechanisms also relieve us from the burden of manual monitoring and submission. With more than $86\,000$ samples used in our experiments (discussed further in Section~\ref{sec:performanceEvaluation}), we could complete the analysis in just $10$ days.

\section{Discussion on Model Updating}

As new malware samples are introduced and evolve daily, model updating is needed to keep the model up to date, thus being able to handle such new samples. However, we believe that this is worth for separate work to develop new methods for model updating such as online learning and dealing with data drift problems. A naive solution is to retrain the teacher model and re-transfer to the student model. The old and new student models are tested on a new test set, and we only deploy the new student if it outperforms the old one by a certain threshold. We will keep this for our future work.

\end{document}